\documentclass[12pt,a4paper]{article}
\usepackage{latexsym}
\usepackage{amsmath}
\usepackage{amsfonts}
\usepackage{amsbsy}
\usepackage{amssymb}
\usepackage{graphicx}
\usepackage{mathrsfs}
\usepackage{epsfig}
\usepackage{bbm}

\def \R{{\mathbb R}}
\def \C{{\mathbb C}}
\def \Z{\mathbb Z}

\def \Diff{\text{Diff }}

\def \A{{\mathfrak A}}

\def \G{\mathcal{G}}

\def \P{\mathcal{P}}

\def \a{\alpha}
\def \b{\beta}

\def \H{\mathcal{H}}
\def \d{\delta}

\def \l{\lambda}
\def \g{\gamma}

\def \s{\sigma}
\def \Sig{\Sigma}

\def \t{\tau}

\def \n{\nu}
\def \m{\mu}

\def \del{\partial}

\def \tr{\text{ tr}}

\def \img{\text{img }}

\newtheorem{Theorem}{Theorem}[section]

\newtheorem{Lemma}{Lemma}[section]
\newtheorem{Definition}{Definition}[section]

\numberwithin{equation}{section}

\begin{document}

\title{Gauge-invariant coherent states for Loop Quantum Gravity I: Abelian gauge groups\\[20pt]}

\author{\it \small Benjamin Bahr, MPI f\"ur Gravitationsphysik, Albert-Einstein
Institut,\\ \it \small Am M\"uhlenberg 1, 14467 Golm, Germany\\[20pt]
\it \small Thomas Thiemann, MPI f\"ur Gravitationsphysik,
Albert-Einstein Institut, \\ \it \small Am M\"uhlenberg 1, 14467
Golm, Germany;\\ \it \small Perimeter Institute for Theoretical
Physics, \\ \it \small 31 Caroline St. N., Waterloo Ontario N2L 2Y5,
Canada}

\maketitle

\abstract{\noindent In this paper we investigate the properties of
gauge-invariant coherent states for Loop Quantum Gravity, for the
gauge group $U(1)$. This is done by projecting the corresponding
complexifier coherent states, which have been applied in numerous
occasions to investigate the semiclassical limit of the kinematical
sector, to the gauge-invariant Hilbert space. This being the first
step to construct \emph{physical} coherent states, we arrive at a
set of gauge-invariant states that approximate well the
gauge-invariant degrees of freedom of abelian LQG. Furthermore,
these states turn out to encode explicit information about the graph
topology, and show the same pleasant peakedness properties known
from the gauge-variant complexifier coherent states.}


\section{Introduction}

\noindent Loop Quantum Gravity (LQG) is a promising candidate for a
theory that aims to combine the principles of quantum mechanics and
general relativity (see \cite{INTRO, ROVELLISBUCH,  INTRO3, ALLMT} and
references therein). The starting point of LQG is the Hamiltonian
formulation of general relativity, choosing Ashtekar-variables as
phase-space coordinates, which casts GR into a $SU(2)$ gauge theory,
leading to the Poisson structure

\begin{eqnarray}
\big\{A_a^I(x)\,,\,A_b^J(y)\big\}\;&=&\;\big\{E_I^a(x)\,,\,E_J^b(y)\big\}\;=\;0\\[5pt]
\big\{A_a^I(x)\,,\,E_J^b(y)\big\}\;&=&\;8\pi
G\b\;\d_{b}^a\,\d_J^I\;\d(x-y).
\end{eqnarray}

\noindent This system could be canonically quantized with the help
of methods well-known from algebraic quantum field theory, which
resulted in a representation of the Poisson-algebra on a Hilbert
space $\H_{kin}$, which carries the kinematical information of
quantum general relativity. One has found recently \cite{LOST} that
this representation is unique up to unitary equivalence if one
demands the space-diffeomorphisms to be unitarily implemented.

While the dynamics of classical general relativity is encoded into a
set of phase-space functions $G_I,\,D_a,\,H$ that are constrained to
vanish, these so-called constraints are, in LQG, promoted to
operators that generate gauge-transformations on the kinematical
Hilbert space $\H_{kin}$. The physical Hilbert space $\H_{phys}$ is
then to be derived as the set of (generalized) vectors being invariant under these
gauge-transformations \cite{HENN-TEITEL}.

\begin{eqnarray}\label{Gl:QuantumConstraints}
\hat G_I|\psi\rangle\;=\;\hat D_a|\psi\rangle\;=\;\hat
H|\psi\rangle\;=\;0.
\end{eqnarray}

\noindent  Although conceptually clear, the actual computation of
$\H_{phys}$ is technically quite difficult. This is due to the fact
that the constraints $\hat G_I,\,\hat D_a\,\hat H$ act quite
non-trivially on $\H_{kin}$. Thus, while the kinematical setting is
understood, the physical states of the theory are not known explicitly. 
It seems that, in its present formulation, LQG is too complicated to
be solved
analytically.\\

While this seems to be discouraging at first, complete solvability
is not something one could have expected from the outset. In fact,
nearly no theory which realistically describes a part of nature is
completely solvable, neither in the quantum, nor in the classical
regime. Rather, having the basic equations of a theory as a starting
point, one has to develop tools for extracting knowledge about its
properties in special cases, reducing the theory to simpler
subsectors, approximate some solutions of the theory, or study its
behavior via numerical methods. Examples for this range from
reducing classical GR to symmetry-reduced situations, which is our
main source of understanding the large-scale structure of our
cosmos, over particle physics, where perturbative quantum field
theory is our access to predict the behavior of elementary
particles, to numerical simulations in ordinary quantum mechanics,
which allow for computations of atomic and molecular spectra,
transition amplitudes or band structures in solid state physics.
Although in all of these fields the fundamental equations are
well-known, their complete solution is elusive, so one has to rely on
approximations and numerics in order to understand the physical
processes described by them. In other cases, such as interacting
Wightman fields on 4D Minkowski space, not a single example is known
to date.
 On the other hand, the perturbation theory
for, say, $SU(N)$-Yang-Mills theory in small couplings is so
effective that many particle physicists even regard the perturbative
expansion in the coupling
parameter as the fundamental theory in itself.\\

With these considerations, it seems quite natural to look for a way
to gain knowledge about the physical content of LQG by approximation
methods. One step into this direction has been done by introducing
the complexifier coherent states.

For ordinary quantum mechanics, the well-known harmonic oscillator
coherent states (HOCS)
\begin{eqnarray}
|z\rangle\;=\;\sum_{n=0}^{\infty}\,\frac{z^n}{\sqrt{n!}}\;|n\rangle
\end{eqnarray}

\noindent are a major tool for performing analytical calculations
and numerical computations. Not only can they be used to approximate
quantum propagators \cite{KECK}, they are also the main tool for
investigating the transition from quantum to classical behaviour, as
well as quantum chaos \cite{KORSCHCHAOS1, KORSCHCHAOS2}. They also
grant access to the numerical treatment of quantum dynamics for
various systems \cite{KLAUDER, VAN-VLECK}, and their generalization
to quantum electrodynamics provides a path to the accurate
description of laser light and quantum optics \cite{GLAUBER}.

The complexifier coherent states (CCS), which have been first
introduced in \cite{HALL1, HALL3}, are a natural generalization of
the HOCS to quantum mechanics on cotangent bundles over arbitrary compact Lie groups, and
the complexifier methods employed to construct these states can also
be transferred to other manifolds as well. Furthermore, for the
special cases of quantum mechanics on the real line $\R$ and the
circle $U(1)$, these states reduce to what has been used as coherent
states for quite some time \cite{KASTRUP, KRP}.

In \cite{CCS}, the complexifier concept has been used to define
complexifier coherent states for LQG. They are states on the
kinematical Hilbert space $\H_{kin}$ and their properties have been
exhibited in \cite{GCS1, GCS2}. It was shown that they mimic the
HOCS in their semiclassical behavior, in the sense that they
describe the quantum system to be close to some point in the
corresponding classical phase-space of general relativity,
minimizing relative fluctuations. Also, they provide a Bargman-Segal
representation of $\H_{kin}$ as holomorphic functions, as well as
approximating well quantum observables that correspond to classical
phase space variables.

This has indicated that these states are a useful tool for examining
the semiclassical limit of LQG. In particular, it has been shown
\cite{TINA1} with the help of the CCS that the constraint operators
for LQG, which are defined on $\H_{kin}$ and generate the dynamics
of the theory, have the correct classical limit. In particular, CCS
that are "concentrated" around a classical solution of GR, are
annihilated by the constraint operators up to orders of $\hbar$.
This indicates that, at least infinitesimally, LQG has classical GR
as semiclassical limit.

On the other hand, since the complexifier coherent states are only
defined on $\H_{kin}$, none of them is really physical in the sense
of the Dirac quantization programme. That is, while they are peaked
on the classical constraint surface, they are not annihilated by the
constraint operators, only approximately.
 Thus,
while being a good tool for examining kinematical properties of LQG,
 it is not clear how well they approximate the
dynamical aspects of quantum general relativity.

To do this, it would be desirable to have coherent states at hand
that satisfy at least some of (\ref{Gl:QuantumConstraints}). We will
pursue the first step on this path in this and the following
article.\\

Some of the constraints (\ref{Gl:QuantumConstraints}) are simpler
than others. In particular, the easiest ones are the Gauss
constraints $\hat G_I$. They are unbounded self-adjoint operators on
$\H_{kin}$ and the gauge-transformations generated by them are well
understood. The set of vectors being invariant under the
Gauss-gauge-transformations ("gauge-invariant" in the following) is
a proper subspace of $\H_{kin}$. This space is well known
\cite{SNF}, and a basis for it is provided by the gauge-invariant
spin network functions, the construction of which involve
intertwiners of the corresponding gauge group $SU(2)$. Thus, the
straightforward way to construct gauge-invariant coherent states
would be to project the CCS to the gauge-invariant Hilbert space. We
will do exactly that in this and the following article.

The gauge transformations correspond to gauging the
$\mathfrak{su}(2)$-valued Ashtekar connection $A_a^I$ and its
canonically conjugate, the electric flux $E_I^a$. Thus, the gauge
group $SU(2)$ is involved, and in fact this group plays a prominent
role in the construction of the whole kinematical Hilbert space
$\H_{kin}$. It is, however, possible to replace $SU(2)$ in this
construction by any compact gauge group $G$, arriving at a different
kinematical Hilbert space $\H_{kin}^{G}$, which would be the arena
for the Hamiltonian formulation of a gauge field theory with gauge
group $G$. Of course, one also has to replace the $\hat G_I$ by the
corresponding gauge generators. Also the constraints $\hat D_a$ and
$\hat H$ can, although nontrivial, be modified to match the new
gauge group. Finally,  the complexifier method is able to
supply corresponding coherent states for each gauge group $G$.

This change of $SU(2)$ into another gauge group has been used
frequently. In \cite{VARA} it has been shown that the quantization
of linearized gravity leads to the LQG framework with $U(1)^3$ as
gauge group. Furthermore, it has been pointed out \cite{QFTCST} that
changing $SU(2)$ for $U(1)^3$ does not change the qualitative
behavior of the theory in the semiclassical limit, and so the
$U(1)^3$-CCS have been
used widely in order to investigate LQG \cite{TINA1}.\\

Before treating the much more complicated case of $G=SU(2)$ in
\cite{GICS-II}, in this paper we will, as a warm-up, consider the
gauge group $G=U(1)$ and the corresponding CCS. The case $G=U(1)^3$
is then simply obtained by a triple tensor product: Not only the
kinematical Hilbert space
\begin{eqnarray}\label{Gl:ThreeTensorProducts}
\H_{kin}^{U(1)^3}\;=\;\H_{kin}^{U(1)}\;\otimes\;\H_{kin}^{U(1)}\;\otimes\;\H_{kin}^{U(1)}
\end{eqnarray}

\noindent has this simple product structure, but also the respective
gauge-invariant subspaces decompose according to
(\ref{Gl:ThreeTensorProducts}). Also, $U(1)^3$-CCS are obtained by
tensoring three $U(1)$-CCS. Due to this simple structure, it is
sufficient for our arguments to consider the gauge-invariant
coherent states in the case of $G=U(1)$, since all the properties
revealed in this article can be carried over straightforwardly to
gauge-invariant coherent states for $G=U(1)^3$.\\

The plan for this paper is as follows: In chapter
\ref{Ch:KinematicalFramework}, we will shortly repeat the basics of
LQG. In particular, the kinematical Hilbert space $\H_{kin}$ for
arbitrary gauge group $G$ is defined, the corresponding set of
constraints that generate the gauge-transformations are described.
In chapter \ref{Ch:TheCCS}, the complexifier coherent states are
defined, where the focus lies on the particular case of $G=U(1)$. A
formula for the inner product between two such states is derived,
which depends purely on the geometry of the complexification of the
gauge group $U(1)^{\C}\simeq\C\backslash\{0\}$. Although this is not
of particular importance in this article, we will find a similar
formula in \cite{GICS-II}, when we come to the case of $G=SU(2)$.
This will hint towards a geometric interpretation of the CCS for
arbitrary gauge groups, and we will comment shortly on this at the
end of \cite{GICS-II}.


In chapter \ref{Ch:GICS} we will apply the projector onto the
gauge-invariant subspace of $\H_{kin}$ to the $U(1)$-complexifier
coherent states. The involved gauge integrals can be carried out by
a special procedure resembling a gauge-fixing. The resulting
gauge-invariant states are then investigated, and their properties
are displayed. In particular, we will show that they describe
semiclassical states peaked at gauge-invariant degrees of freedom.

We will conclude this article with a summary and an outlook to the
sequel paper.

\section{The kinematical setting of
LQG}\label{Ch:KinematicalFramework}

%

\noindent In this section, we will shortly repeat the kinematical
framework of LQG.

Loop Quantum Gravity is a quantization of a Hamiltonian formulation
of classical GR. This is done by introducing an ADM split of
space-time and the introduction of Ashtekar variables \cite{INTRO}.
Thus, GR can be formulated as a constrained SU(2)-gauge theory on a
tree-dimensional manifold $\Sig$, which is regarded as space, and is
taken to be compact. The quantization for noncompact $\Sig$ can also
be carried out, but this requires some more mathematical effort.

On $\Sig$ the Ashtekar $\mathfrak{su}(2)$-connection $A_a^I$ and the
electric flux $E_I^a$ are the dynamical variables. They are
canonically conjugate to each other:

\begin{eqnarray*}
\big\{A_a^I(x)\,,\,A_b^J(y)\big\}\;&=&\;\big\{E_I^a(x)\,,\,E_J^b(y)\big\}\;=\;0\\[5pt]
\big\{A_a^I(x)\,,\,E_J^b(y)\big\}\;&=&\;8\pi
G\b\;\d_{b}^a\,\d_J^I\;\d(x-y).
\end{eqnarray*}

\noindent The fields are not free, but subject to so-called
constraints, which are phase-space functions, i.e. functions of $A$
and $E$. They encode the diffeomorphism-invariance of the theory,
and the Einstein equations. The reduced phase space consists of all
phase space points $A,\,E$ where the constraints vanish. On this
set, the constraints act as gauge transformations, and the set of
gauge orbits is the physical phase space. The set of constraints is
divided into the Gauss constraints $G_I(x)$, the diffeomorphism
constraints $D_a(x)$ and the Hamilton constraints $H(x)$. These
satisfy the Poisson algebra

\begin{eqnarray}\nonumber
\Big\{G(s),\,G(t)\Big\}\;&=&\;G(s\wedge t)\\[5pt]\nonumber
\Big\{G(s),\,D(f)\Big\}\;&=&\;\Big\{G(s),\,H(g)\Big\}\;=\;0\\[5pt]\label{Gl:TheConstraints}
\Big\{D(f),\,D(g)\Big\}\;&=&\;D(\mathcal{L}_fg)\\[5pt]\nonumber
\Big\{D(f),\,H(n)\Big\}\;&=&\;H(\mathcal{L}_fn)\\[5pt]\nonumber
\Big\{H(n),\,H(m)\Big\}\;&=&\;D(g^{ab}(n\,m,_b-m\,n,_b))
\end{eqnarray}

\noindent where $s, t$ are $\mathfrak{su}(2)$-valued functions,
$f,g$ are vector fields on $\Sig$, $n,m$ are scalar functions on
$\Sig$, the smeared constraints are defined by

\begin{eqnarray*}
G(s)\;:=\;\int_{\Sig}G_I(x) s^I(x),\qquad
D(f)\;:=\;\int_{\Sig}D_a(x)\,f^a(x),\qquad
H(n)\;:=\;\int_{\Sig}H(x)\,n(x),
\end{eqnarray*}

\noindent $d$ denotes the exterior derivative on $\Sig$,
$\mathcal{L}$ the Lie derivative, and $\flat$ is the isomorphism
from one-forms to vector fields provided by the metric. It is this
particular occurrence of the metric itself in the Poisson brackets,
which makes the algebra structure notoriously difficult.\\

\subsection{The kinematical Hilbert space}

\noindent The kinematical Hilbert space $\H_{kin}$ of LQG is
computed as a directed limit of Hilbert spaces of functions being
cylindrical over a particular graph embedded in $\Sig$. Consider
$\g$ to be a graph, consisting of finitely many oriented edges
$e_1,\ldots,e_E$ being embedded analytically in $\Sig$, such that
the intersection of two edges is either empty or a common endpoint,
or vertex $v$. For each such graph $\g$ there is a Hilbert space
$H_{\g}$, which consists of all functions being cylindrical over
that particular $\g$. In particular, each edge $E$ of the graph
defines a function from the set of all connections
\begin{eqnarray*}
h_e:\;\mathcal{A}\;\longrightarrow\; SU(2)
\end{eqnarray*}

\noindent by setting $h_e(A)$ being the holonomy of the connection
$A$ along the edge $e$. Symbolically,

\begin{eqnarray*}
h_e(A)\;=\;\P\exp i\int_0^1 dt\;A_a^I(e(t))\frac{\t_I}{2}\,\dot
e^a(t).
\end{eqnarray*}

\noindent A function $f:\mathcal{A}\to\C$ is cylindrical over the
graph $\g$, having $E$ edges $e_1,\ldots e_E$ if there is a function
$\tilde f:SU(2)^E\to\C$ with

\begin{eqnarray}\label{Gl:CorrespondingFunction}
f(A)\;=\;\tilde f\Big(h_{e_1}(A),\,\ldots,\,h_{e_E}(A)\Big).
\end{eqnarray}

\noindent The integration measure in this Hilbert space is just the
Haar measure on $SU(2)^E$, which gives the canonical isomorphism

\begin{eqnarray}\label{Gl:IsomorphismBetweenHGammaAndLTwoOverSU2}
H_{\g}\;\simeq\;L^2\Big(SU(2)^E,\,d\m_H^{\otimes E}\Big).
\end{eqnarray}

\noindent The set of graphs is a partially ordered set. Let
$\g,\,\g'$ be two graphs, then one writes $\g\preceq\g'$, iff there
is a subdivision $\g''$ of $\g'$ by inserting additional vertices
into the edges, such that $\g$ is a subgraph of $\g''$. Note that,
since all graphs consist of analytically embedded edges, this indeed
defines a partially ordering, i.e. for any two graphs $\g_1,\g_2$
there is always a $\g_3$ such that $\g_1\preceq\g_3$ and
$\g_2\preceq\g_3$.

Each function $f_{\g}$ cylindrical over $\g$ determines a
cylindrical function $f_{\g''}$ over $\g''$, simply by defining

\begin{eqnarray}
\tilde f_{\g''}(h_{e_1}(A),\ldots,h_{e_E'}(A))\;:=\;\tilde
f_{\g}(h_{e_{n_1}}(A),\ldots,\,h_{e_{n_E}}(A))
\end{eqnarray}

\noindent where $e_{n_1},\ldots,\,e_{n_E}$ are the edges in $\g''$
belonging to $\g$. Now, every function cylindrical over $\g''$ is
also obviously cylindrical over $\g'$, since $\g''$ is only a
refinement of $\g'$. This procedure defines a unitary map

\begin{eqnarray*}
U_{\g\g'}\;:\;\H_{\g}\;\longrightarrow\;\H_{\g'}.
\end{eqnarray*}

\noindent One can show that for $\g\preceq\g'\preceq\g''$, one has
$U_{\g'\g''}U_{\g\g'}\;=\;U_{\g\g''}$. So, this family of unitary
maps defines a projective limit
\begin{eqnarray}\label{Gl:DirectedLimit}
\H_{kin}\;:=\;\lim_{\longrightarrow}\;\H_{\g},
\end{eqnarray}

\noindent which serves as the kinematical Hilbert space of LQG. Each
$\H_{\g}$ has a canonical isometric embedding $U_{\g}$ into
$\H_{kin}$, which is compatible with the unitary maps $U_{\g\g'}$ in
the following way:

\begin{eqnarray*}
U_{\g\g'}\,U_{\g'}\;=\;U_{\g}\qquad\mbox{for all }\g\preceq\g'.
\end{eqnarray*}

\noindent Due to the definition of the inner product in the
projective limit, for $\psi_{\g}\in\H_{\g}$ and
$\psi_{\g'}\in\H_{\g'}$, where the intersection of $\g$ and $\g'$ is
empty, one has that
\begin{eqnarray*}
\Big\langle U_{\g}\psi_{\g}\Big|U_{\g'}\psi_{\g'}\Big\rangle\;=\;0.
\end{eqnarray*}

\noindent This immediately shows that, since there are uncountably
many graphs with mutual empty intersection in $\Sig$, $\H_{kin}$
cannot be separable. On the other hand, since $\H_{kin}$ is built up
out of the $\H_{\g}$, we can restrict our considerations to an
arbitrary but fixed graph $\g$ for most purposes, dealing only with
the Hilbert space $\H_{\g}$, which is separable.\\

Note that the whole construction carried out here can be done with
an arbitrary compact Lie group $G$. The field $A$ is then a
connection on a $\mathfrak{g}$-bundle and $E$ the corresponding
electric flux, which is canonically conjugate. Also the definition
of the constraints can be adapted to build a theory for arbitrary
gauge groups. This is not only a mathematical toy, but in some
situations, it is in fact useful to replace the gauge group $SU(2)$
by $U(1)^3$, which can be physically justified \cite{GCS2,
VARA, QFTCST}. In particular, we will deal in this article with the
complexifier- and gauge-invariant coherent states for the case of
$G=U(1)$, which will serve as a warm-up example before coming to the
much more difficult (but also more realistic) case of $G=SU(2)$ in
\cite{GICS-II}.

\subsection{Constraint operators and gauge actions}

\noindent In the previous section the kinematical framework for LQG was
presented. In this section, we will shortly discuss the constraint
operators and the gauge actions they induce on $\H_{kin}$.

Rewriting general relativity in a Hamiltonian formulation using the
Ashtekar variables results in the formulation of the Ashtekar
connection $A_a^I(x)$ and the electric flux $E^a_I(x)$, which, in
the quantized theory, become operators on $\H_{kin}$. One cannot
quantize the fields directly, but has to smear them with certain
test functions having support on one-dimensional and two-dimensional
submanifolds of $\Sig$, respectively. See \cite{INTRO} for details.

In the classical theory, the dynamics is encoded in the constraints
(\ref{Gl:TheConstraints}), which in the quantum theory become
operators acting on $\H_{kin}$. The physical Hilbert space is
determined by the condition that (generalized) states are annihilated by the
constraint operators

\begin{eqnarray}\label{Gl:DefinitionOfAPhysicalState}
\hat D_a\;\psi_{phys}\;=\;\hat G_I\;\psi_{phys}\;=\;\hat
H\;\psi_{phys}\;=\;0.
\end{eqnarray}%
%
%

\noindent To implement the Gauss constraints as operators on $\Sig$
is, actually, quite straightforward.
Since the kinematical Hilbert space $\H_{kin}$ can be thought of as
being built up from $\H_{\g}$ for arbitrary graphs $\g\subset\Sig$
by (\ref{Gl:DirectedLimit}), it is sufficient to compute the
gauge-transformation generated by the $\hat G_I$.

 In particular, the similarity between LQG and a lattice gauge
theory on $\g$ is displayed, if one computes the unitary group
generated by the constraints $\hat G_I(x)$, which correspond to
$SU(2)$-gauge transformations of functions on the graph. In
particular, let $k:\Sig\to SU(2)$ be a function and $f$ a
cylindrical function over a graph $\g$ with $E$ edges. The action of
$k$ on $f$ is given by the induced action of $k$ on the
corresponding $\tilde f:SU(2)^E\to \C$ via
(\ref{Gl:CorrespondingFunction}), to be

\begin{eqnarray}\label{Gl:ActionOfGaugeGroup}
\a_{k} \tilde f\;\big(h_{e_1},\ldots,h_{e_E})\;:=\;\tilde
f\;\big(k_{b(e_1)}h_{e_1}k_{f(e_1)}^{-1},\ldots,k_{b(e_E)}h_{e_E}k_{f(e_E)}^{-1}\big),
\end{eqnarray}

\noindent where $b(e_m)$ and $f(e_m)$ are the beginning- and
end-point of the edge $e_m$, and $k_x\in SU(2)$ is the value of the
map $k$ at $x\in\Sig$. So, the gauge transformations act only at the
vertices of a graph.

In particular, one can write down the projector onto the
gauge-invariant Hilbert space for functions in $\H_{\g}$:

\begin{eqnarray}\label{Gl:Projector}
\P
f(h_{e_1},\ldots,h_{e_E})\;&:=&\;\int_{SU(2)^V}d\m_H(k_1,\ldots,k_V)\a_{k_1,\ldots
k_V}\,f(h_{e_1}\ldots,h_{e_E})\\[5pt]\nonumber
&=&\;\int_{SU(2)^V}d\m_H(k_1,\ldots,k_V)f\Big(k_{b(e_1)}h_{e_1}k_{f(e_1)}^{-1},\ldots,k_{b(e_E)}h_{e_E}k_{f(e_E)}^{-1}\Big)
\end{eqnarray}

\noindent Since there are only finitely many vertices on the graph
$\g$, the integral exists and defines a projector
\begin{eqnarray*}
\P:\;\H_{\g}\;\longrightarrow\;\H_{\g}
\end{eqnarray*}

\noindent onto a sub-Hilbert space of $\H_{\g}$. In particular, the
gauge-invariant functions on a graph form a subset of all
cylindrical functions on a graph. The gauge-invariant Hilbert spaces
can be described using intertwiners between irreducible
representations of $SU(2)$, and a basis for the gauge-invariant
Hilbert spaces $\P\H_{\g}$ can be written down in terms of
gauge-invariant spin network functions \cite{SNF}.\\

%

The diffeomorphism constraints $\hat D$ can, however, not be
implemented as operators on $\H_{kin}$ in a straightforward manner.
On the classical side, it can be shown that  the constraint $D(f)$
is the infinitesimal generator of the one-parameter family of
diffeomorphisms defined by the vector field $f$. In particular, a
physical state is one that is invariant under diffeomorphisms, which
simply reflects the invariance of GR under passive (spatial)
diffeomorphisms.

On the quantum side, however, it is straightforward to implement the
action of piecewise analytic diffeomorphisms on $\H_{kin}$: Remember that one
can think of $\H_{kin}$ as consisting of functions
$f:\mathcal{A}\to\C$, which are cylindrical over some graph $\g$.
The space of quantum configurations $\mathcal{A}$, i.e. the space of
(distributional) connections on $\Sig$ carries a natural action of
the diffeomorphism group $\Diff\Sig$. An element $\phi\in\Diff\Sig$
simply acts by $A\to\phi^*A$ on a (distributional) connection $A$.
With this, one can simply define the action of $\Diff\Sig$ on
$\H_{kin}$ by

\begin{eqnarray*}
\a_{\phi}f(A)\;:=\;f(\phi^*A),
\end{eqnarray*}

\noindent where $\phi^*A$ is the pullback of the connection $A$
under the diffeomorphism $\phi$. Note that this definition maps

\begin{eqnarray}\label{Gl:ActionOfDiffeomorphism}
\a_{\phi}\;\H_{\g}\;\longrightarrow\;\H_{\phi(\g)}.
\end{eqnarray}

\noindent Here $\phi(\g)$ is the image of $\g$ under $\phi$. This
shows that one cannot take arbitrary smooth $\phi$, but has to
restrict to analytic diffeomorphisms, since these map a graph
consisting of analytic edges into one consisting again of analytic
edges.

Note that the action (\ref{Gl:ActionOfDiffeomorphism}) is not weakly
continuous in $\phi$, since two graphs can be arbitrary "close" to
each other, but still not intersecting, which means that their
corresponding Hilbert spaces are mutually orthogonal subspaces of
$\H_{kin}$. This fits nicely into the picture, since the notion of
"being close to each other" only has a meaning on manifolds with
metric, and LQG is a quantum theory on a topological manifold only,
since the metric itself is a dynamical object, and not something
given from the outset.\\

\noindent The Hamiltonian constraints $H(n)$ could in fact be
promoted to operators $\hat H(n)$ on $\H_{kin}$ \cite{QSD1}. But,
the solution of this constraint, i.e. determining the set of
(generalized) vectors satisfying $\hat H(n)\psi_{phys}=0$ is still
elusive. Also, since these operators exhibit a highly nontrivial
bracket structure, it is not clear whether they resemble their
classical counterpart (\ref{Gl:TheConstraints}). Moreover, these
operators cannot be defined on the diffeomorphism-invariant Hilbert
space $\H_{\text{diff}}$. To remedy these issues, a modification to
the algebra (\ref{Gl:TheConstraints}) has been proposed, the
so-called master constraint programme. By replacing all $\hat H(n)$
by one operator $\hat M$, one can solve the above issues
\cite{PHOENIX, QSD8}. Still, the solution of this constraint is
quite nontrivial, although some steps into this direction have been
undertaken \cite{TINA1}.\\

\section{Complexifier coherent states}\label{Ch:TheCCS}

\noindent An important question in LQG is whether the theory
contains classical GR in some sort of semiclassical limit
\cite{INTRO,   CCS, TINA1}. The transition from quantum to classical
behavior in the case of, say, a quantum mechanical particle moving
in one dimension can be seen best with the help of the harmonic
oscillator coherent states (HOSZ)
\begin{eqnarray}
|z\rangle\;=\;\sum_{n=0}^{\infty}\,\frac{z^n}{\sqrt{n!}}\;|n\rangle.
\end{eqnarray}

\noindent They can be seen as minimal uncertainty states, or states
that correspond to the system of being in a quantum state close to a
classical phase space point. With these states, one can not only
investigate the transition from quantum to classical behavior of a
system, but one can also try to say something about the dynamics of
the quantum system by considering solutions to the classical
equations of motion.

This has led people to consider, whether states with equally
pleasant properties also exist for LQG. In \cite{CCS}, states in
$\H_{kin}$ have been proposed that have been constructed by the
so-called complexifier method, first brought up in \cite{HALL1,
 HALL3}. They have been investigated in \cite{GCS1, GCS2}, and the
properties of these states seem to make them ideally suited for the
semiclassical analysis of the kinematical sector of LQG
\cite{TINA1}.

The complexifier coherent states are defined for each graph
$\g\subset\Sig$ separately, and each of these Hilbert spaces is, by
(\ref{Gl:IsomorphismBetweenHGammaAndLTwoOverSU2}), a tensor product
of $L^2(SU(2), d\m_H)$-spaces. Also the complexifier coherent states
on $\H_{\g}$ are defined as a tensor product of complexifier
coherent states on $L^2(SU(2), d\m_H)$. In fact, the complexifier
procedure is quite general and works for every compact Lie group
$G$, and is able to define a state on $L^2(G, d\m_H)$. This comes in
handy, since Yang-Mills field theory coupled to gravity can be treated at the kinematical level,
 simply by replacing $SU(2)$ by a compact gauge group $G$ in the whole
construction. There are in fact arguments that, in the semiclassical
limit, the qualitative behavior of calculations in LQG will not
change if one replaces $SU(2)$ by $U(1)^3$. This replacement has
been used widely during the investigation of the semiclassical limit
of LQG \cite{TINA1}. The fact that $U(1)^3$ is abelian is a
tremendous simplification to the calculations.

Thus, in the following we will give the definition of the
complexifier coherent states for arbitrary gauge groups, where the
cases of $G=U(1),\, U(1)^3$ and $SU(2)$ are of ultimate interest for the geometry degrees of freedom of
LQG.

\subsection{General gauge groups}

\noindent Consider quantum mechanics on a compact Lie group $G$,
which is associated to the Hilbert space $L^2(G,\,d\m_H)$, where
$d\m_H$ is the normalized Haar measure on $G$. The classical
configuration space is $G$, and the corresponding phase space is

\begin{eqnarray}
T^*G\;\simeq\;G\times\R^{\dim G}\;\simeq\;G^{\C}.
\end{eqnarray}

\noindent Here, $G^{\C}$ is the complexification of $G$, generated
by the complexification of the Lie algebra of $G$,
$\mathfrak{g}\otimes\C$. The complexifier coherent states are then
defined by
\begin{eqnarray}\label{Gl:DefinitionOfComplexifierCoherntStates}
\psi^t_g(h)\;:=\;\left(e^{\Delta\frac{t}{2}}\;\d_{h'}(h)\right)_{\Big|_{h'\to
g}}.
\end{eqnarray}

\noindent The $\d_{h'}(h)$ is the delta distribution on $G$ with
respect to $d\m_H$, centered around $h'\in G$, $\Delta$ is the
Laplacian operator and $h'\to z$ is the analytic continuation from
$h'\in G$ to $g\in G^{\C}$. The fact that the spectrum of $\Delta$
grows quadratically for large eigenvalues makes sure that the
expression in the brackets is in fact a smooth function on $G$, thus
ensuring that $\psi^t_g\in L^2(G,\,d\m_H)$.

These states are named complexifier coherent states, since, instead
of $-\Delta$, one could have taken any quantization of a phase space
function $C$ (with spectrum bounded from below and spectrum growing
at least as $\l^{1+\epsilon}$, in order for the above expression to
make sense). The function $C$ is called a complexifier, since it
provides an explicit diffeomorphism between $T^*(G)\simeq G^{\C}$,
such that the element $g\in G^{\C}$ actually carries a physical
interpretation as a point in phase space. This diffeomorphism is,
for the complexifier $\hat C=-\Delta$, given by
\begin{eqnarray*}
T^*G\;\simeq\;G\times\R^{\dim G}\;\ni\;(h,\vec
p)\;\longmapsto\;\exp\left(-i\frac{\t_I}{2}p^I\right)h\;\in\;G^{\C}
\end{eqnarray*}

\noindent which is the inverse of the polar decomposition of
elements in $G^{\C}$, while the $\t_I$ are basis elements of
$\mathfrak{g}$. A priori, which complexifier $\hat C$ one chooses is
not fixed. In the context of LQG, one can, given a graph $\g$,
choose a classical function $C$ adapted to this graph, such that its
quantization $\hat C$ is - restricted to $\H_{\g}$ - just the
Laplacian $-\Delta$ on each edge. See \cite{CE} for details and a
discussion of this operator.\\

From (\ref{Gl:DefinitionOfComplexifierCoherntStates}) one can deduce
a more tractable form of the complexifier coherent states given by

\begin{eqnarray}\label{Gl:DefinitionOfComplexifierCoherntStates-2}
\psi^t_g(h)\;=\;\sum_{\pi}e^{-\l_{\pi}}d_{\pi}\,\tr\;\pi(g h^{-1})
\end{eqnarray}

\noindent where the sum runs over all irreducible finite-dimensional
representations $\pi$ of $G$. In the specific case of $G=U(1)$ and
$G=SU(2)$, the states
(\ref{Gl:DefinitionOfComplexifierCoherntStates-2}) have been
investigated \cite{CCS, GCS1, GCS2}, and their properties are known
quite well. In particular, they approximate the quantum operators up
to small fluctuations, the width of which is proportional $\sqrt t$,
which identifies $t$ as the parameter measuring the semiclassicality
scale. For kinematical states in LQG being close to some smooth
space-time, at the scale of say the LHC $t$ is of the order of $l^2_p/(10^{-18}\text{
cm})^2$, i.e. about $10^{-30}$!\\

The states (\ref{Gl:DefinitionOfComplexifierCoherntStates-2}) are
complexifier coherent states for  quantum mechanics on $G$.
Technically, this is equivalent to a graph consisting of one edge.
 For graphs
$\g$ being built of many edges $e_1,\ldots e_E$, one can, since
$L^2(G,d\m_H)^{\otimes E}\;=\;L^2(G^E,d\m_H^{\otimes E})$, simply
construct a state  by taking the tensor product over all edges:

\begin{eqnarray}\label{Gl:TensorProductOfCoherentStates}
\psi^t_{g_1,\ldots, g_E}(h_1,\ldots,
h_E)\;=\;\prod_{m=1}^E\,\psi^t_{g_m}(h_m).
\end{eqnarray}

\noindent Note that this tensor product contains no information
about which edges are connected to each other and which are not.\\

The complexifier coherent states on a graph are labeled by elements
$g_m\in G^{\C}$. In particular, for the cases of interest for LQG,
these spaces are
\begin{eqnarray*}
U(1)^{\C}\;&\simeq&\;\C\backslash\{0\}\\[5pt]
SU(2)^{\C}\;&\simeq&\;SL(2,\C).
\end{eqnarray*}

\noindent As already stated, the complexified groups $G^{\C}$ are
diffeomorphic to the tangent bundle of the groups $T^*G$ themselves.
So, the complexifier coherent states are labeled by elements of the
classical phase space. A state labeled by $g_1,\ldots, g_E$
corresponds to a state being close to the classical phase space
point corresponding to $g_1,\ldots,g_E$. This interpretation is
supported by the fact that - as could be shown for the cases
$G=U(1)$ and $G=SU(2)$ - the expectation values of quantizations of
holonomies and fluxes coincide - up to orders of $\hbar$ - with the
classical holonomies and fluxes determined by the phase space point
corresponding to $g_1,\ldots, g_E$ \cite{GCS2}. Furthermore, the
overlap between two complexifier coherent states is sharply peaked
\cite{GCS2}:

\begin{eqnarray*}
\frac{\Big|\big\langle\psi_{g_1,\ldots,g_E}^t\big|\psi_{h_1,\ldots,h_E}^t\big\rangle\Big|^2}{\big\|\psi_{g_1,\ldots,g_E}^t\big\|^2\;\big\|\psi_{h_1,\ldots,h_E}^t\big\|^2}\;=\;
\left\{\begin{array}{cl}1&\quad g_m=h_m\text{ for all
}m\\\begin{array}{c}\text{ decaying exponentially}\\\text{ as }t\to
0\end{array}& \quad{\rm else}\end{array}\right.
\end{eqnarray*}

\noindent This shows that the complexifier coherent states
(\ref{Gl:DefinitionOfComplexifierCoherntStates-2}) are suitable to
approximate the kinematical operators of LQG quite well. Although
the original LQG has been constructed with $G=SU(2)$, it has been
shown that in the semiclassical regime, the group $SU(2)$ can be
replaced by $U(1)^3$ without changing the qualitative behavior of
expectation values or fluctuations. On the other hand, with this
trick  calculations simplify tremendously, since $U(1)^3$ is an
abelian group. Furthermore, $U(1)^3$ is simply the Cartesian product
of three copies of $U(1)$, which also completely determines the set
of irreducible representations of $U(1)^3$, such that a complexifier
coherent state on $U(1)^3$ is nothing but a product of three states
on $U(1)$:

\begin{eqnarray*}
\psi^t_{(g_1,g_2,g_3)}(h_1,h_2,h_3)\;=\;\psi^t_{g_1}(h_1)\,\psi^t_{g_2}(h_2)\,\psi^t_{g_3}(h_3).
\end{eqnarray*}

\noindent This is, of course, true for any Cartesian product between
- not necessarily distinct - compact Lie groups.

Since the properties of complexifier coherent states on $U(1)^3$ can
be investigated by considering states on $U(1)$, we will work with
the latter from now on.

\subsection{The case of $G=U(1)$}

\noindent In the last section, the general definition of
complexifier coherent states for arbitrary compact Lie groups $G$
has been given. In this section, we will shortly review these states
for the simplest case of $G=U(1)$, since we will work with these
states in the rest of the article.

\noindent From (\ref{Gl:DefinitionOfComplexifierCoherntStates-2}),
we can immediately deduce the explicit form of the complexifier
coherent states, since all irreducible representations of $U(1)$ are
known and one-dimensional:

\begin{eqnarray}\label{Gl:DefinitionOfComplexifierCoherntStatesOnU1}
\psi^t_z(\phi)\;=\;\sum_{n\in\Z}e^{-n^2\frac{t}{2}}\,e^{-in(z-\phi)}
\end{eqnarray}

\noindent for $g=e^{iz}$ and $h=e^{i\phi}$. With the Poisson
summation formula, this expression can be rewritten as
\begin{eqnarray}
\psi_g^t(h)\;=\;\sqrt\frac{2\pi}{t}\;\sum_{n\in\Z}\;e^{-\frac{(z\,-\,\phi\,-\,2\pi
n)^2}{2t}}.
\end{eqnarray}

\noindent The inner product of two of these states is then

\begin{eqnarray}\label{Gl:InnerProductOfTwoU1States}
\big\langle\psi_g^t\big|\psi_{g'}^t\big\rangle\;=\;\sqrt\frac{\pi}{t}\;\sum_{n\in\Z}\;e^{-\frac{(\bar
z\,-\,z'\,-\,2\pi n)^2}{t}}.
\end{eqnarray}

\noindent There is a way to interpret
(\ref{Gl:InnerProductOfTwoU1States}) geometrically. This makes use
of the fact that $G^{\C}\;=\;\C\backslash\{0\}$ comes with a
pseudo-Riemannian metric provided by the Killing form on its Lie
algebra. On arbitrary Lie groups $G$, this metric is denoted, in
components, by

\begin{eqnarray}\label{Gl:ComplexMetric}
h_{IJ}\;=\;-\frac{1}{\dim G}\tr\;\left(g^{-1}\del_Ig\, g^{-1} \del_J
g\right).
\end{eqnarray}

\noindent Choosing the chart $z\to e^{iz}$ on $\C\backslash\{0\}$,
the metric (\ref{Gl:ComplexMetric}) simply takes the form $h=1$.
Note that the geodesics through $1\in\C\backslash\{0\}$ with respect
to this metric are given by

\begin{eqnarray}
t\;\longmapsto\;e^{itz}
\end{eqnarray}

\noindent for some $z\in\C$, which corresponds to the velocity of
the geodesic at $t=0$. Note also that geodesics can be transported
via group multiplication, since the metric is defined via group
translation. In particular, if $\g(t)$ is a geodesic on
$\C\backslash\{0\}$, then $g\g(t)$ is also one for any
$g\in\C\backslash\{0\}$.

With $h$ one can define the complex length-square of a geodesic, or
any other regular curve $\g$ on $\C\backslash\{0\}$, via

\begin{eqnarray}
l^2(\g)\;:=\;\left(\int dt \sqrt{h(\g(t))\dot \g(t)\dot
\g(t)}\right)^2.
\end{eqnarray}

\noindent Note that this gives a well-defined complex number, since
the square of a complex number is defined up to a sign, and this
sign is chosen continuously on the whole curve, which gives a unique
choice since the curve is regular, i.e. its velocity vector vanishes
nowhere. So, the integral is determined up to a sign, the square of
which is then well-defined.

Let $g,\,h\,\in\C\backslash\{0\}$, and
$\g:[0,1]\to\C\backslash\{0\}$ be a geodesic from $g$ to $h$. It is
straightforward to compute that such a geodesic is not unique, but,
for $g=e^{iw}$ and $h=e^{iz}$ (where $z$ and $w$ are determined up
to $2\pi n$ for some $n\in\Z$), is given by

\begin{eqnarray}\label{Gl:Geodesic}
\g(t)\;=\;e^{iw}\,e^{it(z\,-\,w\,-\,2\pi n)}.
\end{eqnarray}

\noindent for any $n\in\Z$. By changing $n$, one ranges through the
set of geodesics from $g$ to $h$. The complex length square of the
(\ref{Gl:Geodesic}) can easily be computed to be

\begin{eqnarray}
l(\g)^2\;=\;(z\,-\,w\,-2\pi n)^2.
\end{eqnarray}

\noindent This shows that one can write the inner product between
two complexifier coherent states as sum over complex lengths of
geodesics:

\begin{eqnarray}
\big\langle\psi^t_g\big|\psi^t_h\big\rangle\;=\;\sum_{\scriptsize\begin{array}{c}\g\text{
geodesic}\\\text{from }g^c \text{ to
}h\end{array}}\;e^{-\frac{l(\g)^2}{t}}
\end{eqnarray}

\noindent with $g^c:=\bar g^{-1}$. \\

Although this seems to be too much effort to rewrite a simple
expression like (\ref{Gl:InnerProductOfTwoU1States}), we will
encounter a similar expression in \cite{GICS-II} for the case of
$SU(2)$-complexifier coherent states. This relates the complexifier
coherent states with the geometry of the corresponding group, which
is given by the Killing metric (\ref{Gl:ComplexMetric}). We will
comment on this at the end of \cite{GICS-II}.



\section{Gauge-invariant coherent states with gauge group
$G=U(1)$}\label{Ch:GICS}

\subsection{The gauge-invariant
sector}\label{Ch:GaugeInvariantSector}


\noindent In the following, we will describe the Hilbert space
invariant under the Gauss gauge transformation group. Since this
gauge transformation group $\G$ leaves every graph invariant, we can
restrict ourselves to the case of one graph, in particular
\begin{eqnarray*}
\P\,\lim_{\longrightarrow}\;H_{\g}\;=\;\lim_{\longrightarrow}\,\P\H_{\g}.
\end{eqnarray*}

\noindent So we can consider the gauge-invariant cylindrical
functions on each graph separately.

The gauge-invariant cylindrical functions on a graph $\g$ with $E$
edges and $V$ vertices can be described in terms of singular
cohomology classes with values in the gauge group. In particular,
every Hilbert space $\H_{\g}$ is canonically isomorphic to an
$L^2$-space:
\begin{eqnarray}
H_{\g}\;\simeq\;L^2\left(G^E,\,d\m_{H}^{\otimes E}\right),
\end{eqnarray}

\noindent where $d\m_H$ is the normalized Haar measure on the
compact Lie group $G$. It is known that the gauge-invariant Hilbert
space is then canonically isomorphic to an $L^2$-space over the
first simplicial cohomology group of $\g$ with values in the gauge
group $G$:
\begin{eqnarray}
\P H_{\g}\;\simeq\;L^2\left(H^1(\g, G),\,d\m\right),
\end{eqnarray}

\noindent with a certain measure $d\m$. For abelian gauge groups
$G$, first cohomology group of $\g$ with values in $G$ is given by
\begin{eqnarray}
H^1(\g, G)\;\simeq\;G^{E-V+1},
\end{eqnarray}

\noindent and $d\m=d\m_H^{\otimes E-V+1}$ is the $E-V+1$-fold tensor
product of the Haar measure on $G$. See appendix
\ref{Ch:Gauge-invariantFunctionsOfU(1)} for a summary of abelian
cohomology groups on graphs and their relation to gauge-invariant
functions. For non-abelian gauge groups $G$ a similar result holds,
while the definition of the first cohomology class requires more
care. This case will be dealt with in \cite{GICS-II}, and we stay
with abelian $G$ in this article.

\subsection{Gauge-invariant coherent states}

\noindent We now come to the main part of this article: The
computation of the gauge-invariant coherent states. We will derive a
closed form for them, revealing the intimate relationship between
the gauge-invariant degrees of freedom and the graph topology. From
the explicit form we will be able to compute the overlap between two
gauge-invariant coherent states, which will allow for an
interpretation as semiclassical states for the gauge-invariant
sector of the theory.\\

The gauge-invariant coherent states are obtained by applying the
gauge projector (\ref{Gl:Projector}) to the complexifier coherent
states on a graph (\ref{Gl:TensorProductOfCoherentStates}),
(\ref{Gl:DefinitionOfComplexifierCoherntStatesOnU1}), i.e.
\begin{eqnarray}\label{Gl:DefinitionGaugeInvariantCoherentState}
\Psi_{[g_1,\ldots,g_E]}^t([h_1,\ldots h_E])\;=\;\P
\psi_{g_1,\ldots,g_E}^t(h_1,\ldots,h_E).
\end{eqnarray}

\noindent It is known that
the set of gauge-invariant functions can be described in terms of
functions on the first cohomology class of the graph. See the
appendix for details. In particular, if the graph has $E$ edges and
$V$ vertices, i.e. the gauge-variant configuration space is
diffeomorphic to $U(1)^E$, then the gauge-invariant configuration
space is diffeomorphic to $U(1)^{E-V+1}$. This might raise the hope
that these states somehow resemble complexifier coherent states on
the gauge-invariant configuration space $U(1)^{E-V+1}$. We will see
that this is not quite true, but near enough.

The fact that the gauge group is abelean is a great simplification:
It allows us to pull back all group multiplications to simple
addition on the algebra, simply due to the fact that
$\exp\,iz\,\exp\,iw=\exp\,i(z+w)$. This will allow us to explicitly
perform the gauge integrals for arbitrary graphs, and obtain a
formula for the gauge-invariant coherent states that only depends on
gauge-invariant combinations of $h_k=\exp\,i\phi_k$ and
$g_k=\exp\,iz_k$, as well as topological information about the
graph, in particular its incidence matrix.

\subsection{Basic graph theory}

In order to be able to deal with the expressions for all graphs, we
start with some basics of graph theory. All the material, as well as
all the proofs, can be found in \cite{GRAPH} and the references
therein.

\begin{Definition}
Let $\g$ be a directed graph with $V$ vertices and $E$ edges. Let
the edges be labeled by numbers $1,\ldots,E$ and the vertices by
numbers $1,\ldots,V$. Then the \emph{incidence matrix}
$\l\in\text{\emph{Mat}}(E\times V,\Z)$ is defined by the following
rule:
\end{Definition}

\begin{eqnarray*}
\l_{kl}\;&:=\;1&\qquad\mbox{if the edge $k$ ends at vertex
$l$}\\[5pt]
\l_{kl}\;&:=\;-1&\qquad\mbox{if the edge $k$ starts at vertex
$l$}\\[5pt]
\l_{kl}\;&:=&\;0\qquad\mbox{else.}
\end{eqnarray*}

\noindent Note in particular that, if edge $k$ starts \emph{and}
ends at vertex $l$, i.e. the edge $k$ is a loop, then $\l_{kl}=0$ as
well. Since either an edge is a loop or starts at one and ends at
some other vertex, every line of the matrix $\l$ is either empty, or
contains exactly one $1$ and one $-1$. With the definition
\begin{eqnarray}\label{Gl:DefintionVonU}
u\;:=\;\left(\begin{array}{c}1\\1\\\vdots\\1\end{array}\right)\;\in\;\R^V,
\end{eqnarray}

\noindent we immediately conclude
\begin{eqnarray}\label{Gl:ULiegtImKernVonLambdaTransponiert}
\l^Tu\;=\;0.
\end{eqnarray}

\begin{Definition}
Let $\g'$ be a graph. If $\g'$ contains no loops, then $\g'$ is said
to be a \emph{tree}. If $\g'\subset\g$ is a subgraph, then $\g'$ is
said to be a \emph{tree in $\g$}. If $\g'\subset\g$ is a subgraph
that meets every vertex of $\g$, then $\g'$ is said to be a
\emph{maximal tree (in $\g$)}.
\end{Definition}

\begin{Lemma}
Every graph $\g$ has a maximal tree as subgraph. Every tree has
$V=E-1$ vertices.
\end{Lemma}

\noindent Maximal trees in graphs are not unique. It is quite easy
to show that every function cylindrical over a graph $\g$ is gauge
equivalent to a function cylindrical over $\g$, which is constant on
the edges corresponding to a maximal tree. This will be used later,
and by the preceding Lemma we immediately conclude that the
number of gauge-invariant degrees of freedoms on a graph with $V$ vertices and
$E$ edges is $E-V+1$ for Abelian gauge theories. This will be seen
explicitly at the end of this section.

The following theorem relates the numbers of different possible
maximal trees to the incidence matrix.

\begin{Theorem}\label{Thm:Kirchhoff} (Kirchhoff)
Let $\g$ be a graph and $\l$ its incidence matrix. Then the
\emph{Kirchoff-matrix} $K:=\l\l^T$ has nonnegative eigenvalues

\begin{eqnarray*}
0\;=\;\m_1\leq\m_2\leq\cdots\leq \m_V.
\end{eqnarray*}

\noindent The lowest eigenvalue is $\m_1=0$, and the degeneracy of
$0$ is the number of connected components of the graph $\g$.
Furthermore, the product of all nonzero eigenvalues
\begin{eqnarray*}
G\;:=\;\frac{1}{V}\prod_{\m_k\neq 0}\m_k
\end{eqnarray*}

\noindent is the number of different maximal trees in $\g$.
\end{Theorem}

With this machinery, we will be able to perform the gauge integral
for arbitrary graphs. This will include some kind of gauge-fixing
procedure, which will make use of a maximal tree.


\subsection{Gauge-variant coherent states and the gauge integral}


\noindent The Abelian nature of the gauge group allows us to pull back
the group multiplication to addition on the Lie algebra. This is why
throughout this chapter we will,  instead of elements $h\in U(1)$,
deal with $\phi\in\R$ by  $h=\exp \,i\phi$, and instead of elements
$g\in\C\backslash\{0\}$, we will work with the corresponding
$z\in\C$ such that $g=\exp\,iz$, always having in mind that $\phi$
and $z$ are only defined modulo $2\pi n$ for $n\in\Z$.

We will denote vectors (of any length) as simple letters
$z,\phi,\tilde\phi,m,\ldots$ and their various components with
indices: $z_k,\phi_k,\tilde\phi_k,\ldots$.
The particular range of the indices will be clear from the context,
but we will still repeat it occasionally.\\

The gauge-variant coherent states on a graph $\g$ with $E$ edges are
simply given by the product
\begin{eqnarray}
\psi_{z}^t(\phi)\;=\;\prod_{k=1}^E\sum_{m_k\in\Z}e^{-m_k^2\frac{t}{2}}\,e^{im_k(z_k-\phi_k)}
\end{eqnarray}

\noindent where $z_k=\phi_k-ip_k,\;k=1,\ldots,E$ is labeling the
points in phase space where the coherent states are peaked. With the
Poisson summation formula one can rewrite this as

\begin{eqnarray}\label{Gl:EichvarianterKohaerenterZustandAufNemGraphen}
\psi_{z}^t(\phi)\;=\;\sqrt{\frac{2\pi}{t}}^E\;\sum_{m_1,\ldots,m_E\in\Z}\;\exp\left({-\sum_{k=1}^E\frac{(z_k-\phi_k-2\pi
m_k)^2}{2t}}\right)
\end{eqnarray}

\noindent We will now perform the gauge integral

\begin{eqnarray}\label{Gl:AusgangsFormel}
\Psi_{[z]}^t(\phi)\;&=&\;\int_Gd\m_H(\tilde
\phi)\;\psi_{\a_{\tilde\phi}z}(\phi)\\[5pt]\nonumber
&=&\;\sqrt{\frac{2\pi}{t}}^E\int_{[0,2\pi]^V}\frac{d\tilde\phi_1}{2\pi}\cdot\ldots\cdot\frac{d\tilde\phi_V}{2\pi}\;
\sum_{m_1,\ldots,m_E\in\Z}\;\exp\left({-\sum_{k=1}^E\frac{(A_k+\l_{ka}\tilde\phi_a-2\pi
m_k)^2}{2t}}\right)
\end{eqnarray}

\noindent with $A=z-\phi$, and where $\l_{ka}$ are the components of
the transpose $\l^T$ of the incidence matrix.

In what follows, we will use the symmetries of this expression,
together with a gauge-fixing procedure, to separate the gauge
degrees of freedom from the gauge-invariant ones. The integrals will
then be performable analytically, and the resulting expression can
then be interpreted as states being peaked on gauge-invariant
quantities.\\

\noindent To simplify the notation, we will assume, without loss of
generality, that $\g$ is connected. Furthermore choose, once and for
all, a maximal tree $\t\subset\g$. Choose the numeration of vertices
 and edges  of $\g$ according to the
following scheme:

Start with the maximal tree $\t$. The tree consists of $V$ vertices
and $V-1$ edges. Call a vertex that has only one outgoing edge (in
$\t$, not necessarily in $\g$) an outer end of $\t$. Remove one
outer end and the corresponding edge from $\t$ and obtain a smaller
subgraph $\t^1\subset\g$, which is also a tree. Label the removed
vertex with the number $1$, and do so with the removed edge as well.
So this gives you $v_1$ and $e_1$. From $\t^1$, remove an outer end
and the corresponding edge, and label them $v_2$ and $e_2$, and
obtain a yet smaller tree $\t^2\subset \t^1\subset \t\subset\g$.
Repeat this process until $\t$ has been reduced to $\t^{
(V-1)}$, which is a point. This way, one has obtained
$v_1,\ldots,v_{V-1}$ and $e_1,\ldots,e_{E-1}$. Call the last,
remaining vertex $v_V$. Label the edges that do not belong to $\t$
by $e_V,e_{V+1},\ldots,e_E$ in any order.

Choosing the numeration of the vertices and the edges in the above
manner will help us in rewriting the expression
(\ref{Gl:AusgangsFormel}). First we note that the first $V-1$ edges
and the first $V$ vertices constitute the tree, the last $E-V+1$
edges constitute what is not the tree in $\g$. Furthermore, with
this numeration, the edge $e_k$ is starting or ending at vertex
$v_k$ for $k=1,\ldots,V-1$. In particular, the diagonal elements of
the incidence matrix are all (except maybe the last one) nonzero:
$\l_{kk}\neq 0$ for $k=1,\ldots,V-1$.

\begin{Definition}
Let $\g$ be a graph, with vertices $v_1,\ldots v_V$ and edges
$e_1,\ldots,e_E$. Between two vertices $v_k$ and $v_l$ there is a
unique path in $\t$, since a tree contains no loops. Call $v_k$
being \emph{before} $v_l$, if this path includes $e_k$, otherwise
call $v_k$ being \emph{after} $v_l$.
\end{Definition}

\noindent Note that a vertex cannot be both before and after another
vertex, but two vertices can both be before or both be after each
other.

The numeration we have chosen has the following consequence: For
each vertex $v_k$ one has that for all $v_l$ such that $v_k$ is
after $v_l$, that $l\leq k$. The converse need not be true. Note
further that every vertex is before itself, by this definition.
Also, since $e_V$ is not an edge of the graph, it does not even have
to be touching $v_V$. So, the question of whether $v_V$ is before or
after any other vertex makes no sense in this definition (But note
that it does make sense to ask whether any vertex is before or after
$v_V$).


We now rewrite formula (\ref{Gl:AusgangsFormel}), by replacing the
integrals over $[0,2\pi]$ by integrals over $\R$. We do this
inductively over the vertices from $v_1$ to $v_{V-1}$. Consider the
$E$ terms constituting the sum in the exponent in

\begin{eqnarray*}
\Psi_{[z]}^t(\phi)\;&=&\;\int_Gd\m_H(\tilde
\phi)\;\psi_{\a_{\tilde\phi}z}(\phi)\\[5pt]\nonumber
&=&\;\sqrt{\frac{2\pi}{t}}^E\int_{[0,2\pi]^V}\frac{d\tilde\phi_1}{2\pi}\cdot\ldots\cdot\frac{d\tilde\phi_V}{2\pi}\;
\sum_{m_1,\ldots,m_E\in\Z}\;\exp\left({-\sum_{k=1}^E\frac{(A_k+\l_{ka}\tilde\phi_a-2\pi
m_k)^2}{2t}}\right).
\end{eqnarray*}

\noindent In some of them $\tilde\phi_1$ appears, in some of them it
does not, precisely if either $\l_{k1}\neq 0$ or $\l_{k1}=0$. Note
that $\tilde\phi_1$ definitely appears in the first term, by the
above considerations. If $\tilde\phi_1$ appears in the $k$-th term
other than $k=1$, shift the infinite sum over $m_k$ by
$m_k+\l_{11}\l_{k1}m_1$. The result of this is that, since
$\l_{k1}^2=\l_{11}^2=1$ for these $k$, after this shift
$\tilde\phi_1$ appears always in the combination
$\l_{11}\tilde\phi_1-2\pi m_1$ in all the factors. Now we can employ
the formula
\begin{eqnarray}\label{Gl:SuperFormelDieSummenWegmacht}
\int_{[0,2\pi]}\frac{d\tilde\phi}{2\pi}\,\sum_{m\in\Z}\;f(\tilde\phi
\pm 2\pi m)\;=\;\frac{1}{2\pi}\int_{\R}d\tilde\phi\;f(\tilde\phi)
\end{eqnarray}

\noindent and, regardless of whether $\l_{11}=+1$ or $\l_{11}=-1$,
have
\begin{eqnarray*}
(\ref{Gl:AusgangsFormel})\;&=&\;\sqrt{\frac{2\pi}{t}}^E\int_{\R}\frac{d\tilde\phi_1}{2\pi}\int_{[0,2\pi]^{V-1}}
\frac{d\tilde\phi_2}{2\pi}\cdot\ldots\cdot\frac{d\tilde\phi_V}{2\pi}\;\\[5pt]
&&\qquad\qquad\qquad\times\sum_{m_2,\ldots,m_E\in\Z}\;\exp\left({-\frac{(A_1+\l_{1a}\tilde\phi_a)^2}{2t}-\sum_{k=2}^E\frac{(A_k+\l_{ka}\tilde\phi_a-2\pi
m_k)^2}{2t}}\right).
\end{eqnarray*}

\noindent This being the beginning of the induction, we now describe
the induction step from $l$ to $l+1$ by the following technical
lemma. By this we will be able to extend all integration ranges over
all of $\R$, instead of finite intervals, which will turn out to be
very useful.

\begin{Lemma}\label{Lem:InduktionsSchrittBeimSummenverschwindenlassen}
\noindent Let $\g$ be a graph with $V$ vertices, $E$ edges, and $\l$
be its incidence matrix. Let $A\in\C^E$ and $t>0$, then we have, for
$1\leq l\leq V-1$:
\begin{eqnarray*}
&&\sqrt{\frac{2\pi}{t}}^E\int_{\R^{l-1}}\frac{d\tilde\phi_1}{2\pi}\cdot\ldots\cdot\frac{d\tilde\phi_{l-1}}{2\pi}\int_{[0,2\pi]^{V-l+1}}
\frac{d\tilde\phi_{l}}{2\pi}\cdot\ldots\cdot\frac{d\tilde\phi_V}{2\pi}\;\\[5pt]
&&\qquad\qquad\qquad\times\sum_{m_{l},\ldots,m_E\in\Z}\;\exp\left({-\sum_{k=1}^{l-1}\frac{(A_k+\l_{ka}\tilde\phi_a)^2}{2t}-\sum_{k={l}}^E\frac{(A_k+\l_{ka}\tilde\phi_a-2\pi
m_k)^2}{2t}}\right)\\[5pt]
&=&\sqrt{\frac{2\pi}{t}}^E\int_{\R^{l}}\frac{d\tilde\phi_1}{2\pi}\cdot\ldots\cdot\frac{d\tilde\phi_{l}}{2\pi}\int_{[0,2\pi]^{V-l}}
\frac{d\tilde\phi_{l+1}}{2\pi}\cdot\ldots\cdot\frac{d\tilde\phi_V}{2\pi}\;\\[5pt]
&&\qquad\qquad\qquad\times\sum_{m_{l+1},\ldots,m_E\in\Z}\;\exp\left({-\sum_{k=1}^{l}\frac{(A_k+\l_{ka}\tilde\phi_a)^2}{2t}-\sum_{k={l+1}}^E\frac{(A_k+\l_{ka}\tilde\phi_a-2\pi
m_k)^2}{2t}}\right).
\end{eqnarray*}
\end{Lemma}

\noindent\textbf{Proof:} Note that we just proved the formula for
$l=1$. In the proof for arbitrary $1\leq l\leq V-1$ we will use the
notion of vertices being before and after one another.

Consider all vertices $v_k$ being after $v_l$, other than $v_l$
itself. By construction, for all such $k$, we have $k<l$, so by
induction hypothesis, the integration over all these $v_k$ runs over
all of $\R$, not over just the interval $[0,2\pi]$ any more.
Consequently, the sum over these $m_k$ is not appearing any longer.
So we can shift the integration range by $+2\pi\l_{ll}m_l$.

%

This will affect the terms in the first sum in
\begin{eqnarray}\label{Gl:DieBeidenSummen}
\exp\left(-\sum_{k=1}^{l-1}\frac{(A_k+\l_{ka}\tilde\phi_a)^2}{2t}-\sum_{k={l}}^E\frac{(A_k+\l_{ka}\tilde\phi_a-2\pi
m_k)^2}{2t}\right)
\end{eqnarray}

\noindent in the following way: Let $k<l$. The edge $e_k$ then belongs
to the tree $\t$, and thus $v_l$ is either after both vertices $e_k$
touches, or before both vertices. If $v_l$ is before both, the term
does not change at all, since the two $\tilde\phi_a$ in it are not
shifted. If $v_l$ is after both and is not itself one of the two
vertices, then the term gets changed by
\begin{eqnarray*}
(A_k+\l_{ka}\tilde\phi_a)^2\;\longrightarrow\;(A_k+\l_{ka}\tilde\phi_a\;\pm2\pi\l_{ll}m_l\;\mp
2\pi\l_{ll}m_l)^2\;=\;(A_k+\l_{ka}\tilde\phi_a)^2,
\end{eqnarray*}

\noindent since the two $\tilde\phi_a$ in a term always appear with
opposite sign. So these terms do not change, too. If $v_l$ is after
both vertices that touch $e_k$ and is itself one of it (i.e. $e_k$
is an edge adjacent to $e_l$, linked by $v_l$), then the
corresponding term changes by

\begin{eqnarray*}
(A_k+\l_{ka}\tilde\phi_a)^2\;&=&\;
(A_k+\l_{kl}\tilde\phi_l+\l_{kk}\tilde\phi_k)^2\;=\;(A_k+\l_{kk}(\tilde\phi_k-\tilde\phi_l))^2\\[5pt]
&&\longrightarrow\;(A_k+\l_{kk}(\tilde\phi_k-\tilde\phi_l+2\pi\l_{ll}m_l))^2,
\end{eqnarray*}

\noindent where $\l_{kl}=-\l_{ll}$ and $\l_{ll}^2=1$ have been used.

So, after this shift, in all terms in the first sum in
(\ref{Gl:DieBeidenSummen}) $\tilde\phi_l$ has been replaced by
$\tilde\phi_l-2\pi\l_{ll}m_l$. The first term of the second sum
reads

\begin{eqnarray*}
(A_l+\l_{ll}(\tilde\phi_l-\tilde\phi_a)-2\pi
m_l)^2\;=\;(A_l-\l_{ll}\tilde\phi_a+\l_{ll}(\tilde\phi_l-2\pi\l_{ll}
m_l))^2,
\end{eqnarray*}

\noindent where $v_a$ is the other vertex touching $e_l$, apart from
$v_l$. So also in this term $\tilde\phi_l$ and $m_l$ appear in the
combination $\tilde\phi_l-2\pi\l_{ll}m_l$.

The terms $k=l+1$ till $k=E-V+1$ remain unchanged, since they all
correspond to edges that lie between vertices $v_a$ such that $v_l$
is before both $v_a$, and these $\tilde\phi_a$ are hence not
shifted.

The terms $k=E-V+2$ till $k=E$ in (\ref{Gl:DieBeidenSummen}), on the
other hand, correspond to edges that lie between two vertices such
that $v_l$ could be before the one and after the other. This is due
to the fact that these edges do not belong to the maximal tree $T$
any longer. So in these terms, a shift by $\pm2\pi\l_{ll}m_l$ could
have occurred by the shift of integration range. But in all these
terms, there is still a term $-2\pi m_k$ present, and the sum over
these $m_k$ is still performed. So, by appropriate shift of these
summations, similar to the ones performed in the induction start,
one can subsequently produce or erase terms of the form
$\pm2\pi\l_{ll}m_l$ in all of the terms corresponding to $k=E-V+2$
till $k=E$. Since there are enough summations left, one has enough
freedom to produce a $\pm2\pi\l_{ll}m_l$, where $\tilde\phi_l$ is
present, or erase all terms with $m_l$, where $\tilde\phi_l$ is not
present.\\

Thus, in the end, we again have a function only depending on
$\tilde\phi_l-2\pi\l_{ll}m_l$, and thus we can again apply formula
(\ref{Gl:SuperFormelDieSummenWegmacht}), and, regardless of the sign
of $\l_{ll}$, erase the infinite sum over $\m_l$, obtaining an
integration range of $\tilde\phi_l$ over all of $\R$:

\begin{eqnarray*}
&&\sqrt{\frac{2\pi}{t}}^E\int_{\R^{l-1}}\frac{d\tilde\phi_1}{2\pi}\cdot\ldots\cdot\frac{d\tilde\phi_{l-1}}{2\pi}\int_{[0,2\pi]^{V-l+1}}
\frac{d\tilde\phi_{l}}{2\pi}\cdot\ldots\cdot\frac{d\tilde\phi_V}{2\pi}\;\\[5pt]
&&\qquad\qquad\qquad\times\sum_{m_{l},\ldots,m_E\in\Z}\;\exp\left({-\sum_{k=1}^{l-1}\frac{(A_k+\l_{ka}\tilde\phi_a)^2}{2t}-\sum_{k={l}}^E\frac{(A_k+\l_{ka}\tilde\phi_a-2\pi
m_k)^2}{2t}}\right)\\[5pt]
&=&\sqrt{\frac{2\pi}{t}}^E\int_{\R^{l}}\frac{d\tilde\phi_1}{2\pi}\cdot\ldots\cdot\frac{d\tilde\phi_{l}}{2\pi}\int_{[0,2\pi]^{V-l}}
\frac{d\tilde\phi_{l+1}}{2\pi}\cdot\ldots\cdot\frac{d\tilde\phi_V}{2\pi}\;\\[5pt]
&&\qquad\qquad\qquad\times\sum_{m_{l+1},\ldots,m_E\in\Z}\;\exp\left({-\sum_{k=1}^{l}\frac{(A_k+\l_{ka}\tilde\phi_a)^2}{2t}-\sum_{k={l+1}}^E\frac{(A_k+\l_{ka}\tilde\phi_a-2\pi
m_k)^2}{2t}}\right).
\end{eqnarray*}

\noindent This was the claim of the Lemma.\\

An immediate corollary of Lemma
\ref{Lem:InduktionsSchrittBeimSummenverschwindenlassen} is that

\begin{eqnarray}\nonumber
&&\sqrt{\frac{2\pi}{t}}^E\int_{[0,2\pi]^V}\frac{d\tilde\phi_1}{2\pi}\cdot\ldots\cdot\frac{d\tilde\phi_V}{2\pi}\;
\sum_{m_1,\ldots,m_E\in\Z}\;\exp\left({-\sum_{k=1}^E\frac{(A_k+\l_{ka}\tilde\phi_a-2\pi
m_k)^2}{2t}}\right)\\[5pt]\label{Gl:AlleIntegrationenGeshiftet}
&=&\;\sqrt{\frac{2\pi}{t}}^E\int_{\R^{V-1}}\frac{d\tilde\phi_1}{2\pi}\cdot\ldots\cdot\frac{d\tilde\phi_{V-1}}{2\pi}\int_0^{2\pi}
\frac{d\tilde\phi_{V}}{2\pi}\\[5pt]\nonumber
&&\qquad\qquad\qquad\times\sum_{m_{V},\ldots,m_E\in\Z}\;\exp\left({-\sum_{k=1}^{V-1}\frac{(A_k+\l_{ka}\tilde\phi_a)^2}{2t}-
\sum_{k={V}}^E\frac{(A_k+\l_{ka}\tilde\phi_a-2\pi
m_k)^2}{2t}}\right).
\end{eqnarray}

\noindent Note that one cannot perform the induction step with the
integration over $\tilde\phi_V$ as well. The reason for this is that
for the induction step it is crucial that it does not make sense to
define whether $v_V$ is before or after any other vertex, since
$e_V$ does not belong to the maximal tree $\t$, in fact it does not
even need to touch $v_V$. In particular, the integrand in
(\ref{Gl:AlleIntegrationenGeshiftet}) does not depend on
$\tilde\phi_V$ at all! To see this, one only needs to shift all
integrations $\tilde\phi_1,\ldots,\tilde\phi_{V-1}$ by
$+\tilde\phi_V$. In all terms, the integration variables appear in
the combination $\tilde\phi_a-\tilde\phi_b$ for any two different
$a,b=1,\ldots,V$. So either $a$ and $b$ are both not $V$, then
nothing changes by this shift of integration, or one of $a$ or $b$
is equal to $V$. In this case the shift of the other one cancels the
$\tilde\phi_V$, since both $\tilde\phi_a$ and $\tilde\phi_b$ appear
with opposite sign. So, after this shift, $\tilde\phi_V$ occurs
nowhere in the formula any more. Thus, we can perform the
integration over $\tilde\phi_V$ trivially and obtain
\begin{eqnarray}\label{Gl:FormelMitZuNullGesetzemPhiTildeVau}
(\ref{Gl:AusgangsFormel})\;=\;\sqrt{\frac{2\pi}{t}}^E\int_{\R^{V-1}}\frac{d\tilde\phi_1}{2\pi}\cdot\ldots\cdot\frac{d\tilde\phi_{V-1}}{2\pi}\;
\sum_{m_{V},\ldots,m_E\in\Z}\;\exp\left(-\sum_{k=1}^{E}\frac{(\tilde
A_k+\l_{ka}\tilde\phi_a)^2}{2t}\right)_{\Bigg|_{\tilde\phi_V=0}}
\end{eqnarray}

\noindent where
\begin{eqnarray}
\tilde  A_k\;:=\left\{\begin{array}{ll}A_k&1\leq k\leq V-1\\A_k-2\pi
m_k\qquad & V\leq k\leq E\end{array}\right..
\end{eqnarray}

\noindent To proceed, note that, since in every term in
(\ref{Gl:FormelMitZuNullGesetzemPhiTildeVau}) the $\tilde\phi_a$
appear as pairs with opposite sign, the integrand is invariant under
a simultaneous shift of all variables:
$\tilde\phi_a\to\tilde\phi_a+c$.  We use this fact to rewrite
(\ref{Gl:FormelMitZuNullGesetzemPhiTildeVau}), by using the
following technical Lemma

\begin{Lemma}\label{Lem:WieKommtDieDeltaFunktionInDieFlasche}
Let $f:\R^n\to \C$ be a function with the symmetry
\begin{eqnarray*}
f(x_1+c,\ldots,x_n+c)\;=\;f(x_1,\ldots,x_n)\qquad\mbox{for all
}c\in\R
\end{eqnarray*}

\noindent such that $x_1,\ldots,x_{n-1}\to f(x_1,\ldots,x_{n-1},0)$
is integrable. Then
\begin{eqnarray}
\int_{\R^{n-1}}d^{n-1}x\;f(x_1,\ldots,x_{n-1},0)\;=\;n\int_{\R^n}d^nx\,\d(x_1+\cdots +x_n)\,f(x_1,\ldots,x_n).
\end{eqnarray}
\end{Lemma}

\noindent\textbf{Proof: } The proof is elementary. Write

\begin{eqnarray*}
&&\int_{\R^{n-1}}dx_1,\ldots dx_{n-1}\;f(x_1,\ldots,x_{n-1},0)\\[5pt]
\;&=&\;\int_{\R^{n-1}}dx_1,\ldots
dx_{n-1}\;f\left(x_1-\frac{\sum_{k=1}^{n-1}
x_k}{n},\ldots,x_{n-1}-\frac{\sum_{k=1}^{n-1}
x_k}{n},-\frac{\sum_{k=1}^{n-1} x_k}{n}\right)\\[5pt]
\;&=&\;\int_{\R^n}dx_1,\ldots
dx_{n}\;f\left(x_1-\frac{\sum_{k=1}^{n-1}
x_k}{n},\ldots,x_{n-1}-\frac{\sum_{k=1}^{n-1}
x_k}{n},\,x_n\right)\,\d\left(x_n+\frac{\sum_{k=1}^{n-1}x_k}{n}\right)
\end{eqnarray*}

\noindent Now perform a coordinate transformation
\begin{eqnarray*}
&&\tilde x_k\;:=\;x_k\,-\frac{\sum_{k=1}^{n-1}
x_k}{n},\;\qquad\mbox{
for }k=1,\ldots ,n-1\\[5pt]
&&\tilde x_n\;:=\;x_n.
\end{eqnarray*}

\noindent We have

\begin{eqnarray*}
\sum_{n=1}^{n-1}\tilde x_k\;=\;\frac{\sum_{k=1}^{n-1}x_k}{n}
\end{eqnarray*}

and get
\begin{eqnarray}\nonumber
&&\int_{\R^{n-1}}dx_1,\ldots dx_{n-1}\;f(x_1,\ldots,x_{n-1},0)\\[5pt]\label{Gl:JetztFehltNurNochDieJacobimatrix}
\;&=&\;\frac{1}{J}\int_{\R^n}d^n\tilde x\;f\left(\tilde
x_1,\ldots,\tilde x_{n-1},\,\tilde x_n\right)\;\d(\tilde
x_1+\ldots+\tilde x_n).
\end{eqnarray}

\noindent Here $J=\det{(\del \tilde x_k/\del  x_l)}$ is the Jacobian
matrix of the coordinate transform. It is given by
\begin{eqnarray*}
J\;=\;\det\left[1\,-\,\frac{1}{n}\left(\begin{array}{ccccc}1&1&\cdots&1&0\\1&1&\cdots&1&0\\
\vdots & \vdots&\ddots&\vdots&\vdots\\1&1&\cdots&1&0\\0&0&\cdots
&0&0
\end{array}\right)\right],
\end{eqnarray*}

\noindent the determinant of which can easily computed to be
$J=\frac{1}{n}$. Thus, with
(\ref{Gl:JetztFehltNurNochDieJacobimatrix}), the statement is
proven.\\

We continue our analysis of the gauge-invariant overlap by using
Lemma (\ref{Lem:WieKommtDieDeltaFunktionInDieFlasche}) to rewrite
(\ref{Gl:FormelMitZuNullGesetzemPhiTildeVau}) to obtain

%
%
%
\begin{eqnarray*}
(\ref{Gl:AusgangsFormel})\;=\;V\sqrt{\frac{2\pi}{t}}^E\int_{\R^{V}}\frac{d\tilde\phi_1\ldots
d\tilde\phi_{V}}{(2\pi)^{V-1}}\;\d\left(\sum_{a=1}^V\tilde\phi_a\right)
\sum_{m_{V},\ldots,m_E\in\Z}\;\exp\left(-\sum_{k=1}^{E}\frac{(\tilde
A_k+\l_{ka}\tilde\phi_a)^2}{2t}\right).
\end{eqnarray*}

\noindent Now we split the integrations over the $\tilde\phi_a$ from
the $\tilde A_k$, in order to perform the integration. Because we
are integrating over $\R^V$ and the integrand is holomorphic, we can
now shift the $\tilde\phi_a$ also by complex amounts. This is
necessary, since the $\tilde A_k$ are generically complex. A generic
shift of the $\tilde\phi_a$ by complex numbers $z_a$ looks like
\begin{eqnarray}\nonumber
(\ref{Gl:AusgangsFormel})\;&=&\;V\sqrt{\frac{2\pi}{t}}^E\int_{\R^{V}}\frac{d\tilde\phi_1\ldots
d\tilde\phi_{V}}{(2\pi)^{V-1}}\;\d\left(\sum_{a=1}^V(\tilde\phi_a+z_a)\right)\\[5pt]\nonumber
&&\quad\times
\sum_{m_{V},\ldots,m_E\in\Z}\;\exp\left(-\sum_{k=1}^{E}\frac{(\tilde
A_k+\l_{ka}\tilde\phi_a+\l_{ka}z_a)^2}{2t}\right)\\[5pt]\nonumber
\;&=&\;V\sqrt{\frac{2\pi}{t}}^E\int_{\R^{V}}\frac{d\tilde\phi_1\ldots
d\tilde\phi_{V}}{(2\pi)^{V-1}}\;\d\left(\sum_{a=1}^V(\tilde\phi_a+z_a)\right)\\[5pt]\nonumber
&&\quad\times\sum_{m_{V},\ldots,m_E\in\Z}\;\exp\left[-\sum_{k=1}^{E}\left(\frac{(\l_{ka}\tilde\phi_a)^2}{2t}
\,+\,\frac{\l_{ka}\tilde\phi_a(\l_{ka}z_a+\tilde
A_k)}{t}\,+\,\frac{(\l_{ka}z_a+\tilde
A_k)^2}{2t}\right)\right]\\[5pt]\label{Gl:AllesMitVektorenAusgedrueckt}
\;&=&\;V\sqrt{\frac{2\pi}{t}}^E\int_{\R^{V}}\frac{d\tilde\phi_1\ldots
d\tilde\phi_{V}}{(2\pi)^{V-1}}\;\d\left(u^T\tilde\phi+u^Tz\right)\\[5pt]\nonumber
&&\quad\times\sum_{m_{V},\ldots,m_E\in\Z}\;\exp\left(-\frac{\tilde\phi^T\l\l^T\tilde\phi}{2t}
\,-\,\frac{\tilde\phi^T\l(\l^Tz+\tilde
A)}{t}\,-\,\frac{(\l^Tz+\tilde A)^T(\l^Tz+\tilde A)}{2t}\right).
\end{eqnarray}

\noindent In (\ref{Gl:AllesMitVektorenAusgedrueckt}) we have
expressed all variables in terms of vectors and matrices, since this
will simplify the handling of the expressions a lot. The vectors
$u$, $\tilde\phi$, $z$ have length $V$, the vector $\tilde A$ has
length $E$, and $\l$ is the $V\times E$ incidence matrix. The vector
$u$ is given by (\ref{Gl:DefintionVonU}). The ${}^T$ means
transpose.

The following Lemma will help us to simplify this formula.

\begin{Lemma}\label{Lem:LoesungenVonGleichungssystemen}
Let $\l$ be the $V\times E$ incidence matrix of a connected graph
$\g$ with $E$ edges and $V$ vertices, and $u=(1\,1\,\cdots\,1)^T$
the vector of length $V$ containing only ones. For any vector
$\tilde A\in \C^E$ the set of equations
\begin{eqnarray*}
\l(\l^Tz+\tilde A)\;&=&\;0\\[5pt]
u^Tz\;&=&\;0
\end{eqnarray*}

\noindent has exactly one solution in $\C^V$.
\end{Lemma}

\noindent\textbf{Proof:} Rewrite the first of these equations as
\begin{eqnarray*}
\l\l^T\,z\;=\;-\l\tilde A.
\end{eqnarray*}

\noindent Because of (\ref{Gl:ULiegtImKernVonLambdaTransponiert}),
$-\l\tilde A$ lies in the orthogonal complement of $u$: $-\l\tilde
A\in\{u\}^{\perp}$. Since the graph $\g$ is connected, by
Kirchhoff's theorem (\ref{Thm:Kirchhoff}) the Kirchhoff-matrix
$\l\l^T$ is positive definite on $\{u\}^{\perp}$, hence invertible
on this $V-1$-dimensional subspace of $\C^V$. Define the $V\times
V$-matrix $\s$ to be the inverse of $\l\l^T$ on $\{u\}^{\perp}$, and
zero on $u$:
\begin{eqnarray*}
\s(\l\l^T)v\;=\;(\l\l^T)\s v\;&=&\;v\qquad\mbox{for all }u^Tv=0\\[5pt]
\s u\;&=&\;0.
\end{eqnarray*}

\noindent So, the set of solutions of $\l\l^Tz=-\l\tilde A$ is given
by
\begin{eqnarray}
z\;=\;-\s\l\tilde A\;+\;\a u\;\qquad\a\in\C.
\end{eqnarray}

\noindent By the definition of $\s$, this means that
\begin{eqnarray}
z\;=\;-\s\l\tilde A
\end{eqnarray}

\noindent is the unique solution of both equations, which proves the
Lemma.\\

\begin{Lemma}\label{Lem:DasIstJaEinProjektor!}
With the conditions of Lemma
\ref{Lem:LoesungenVonGleichungssystemen}, let $z$ be the unique
solution of $\l(\l^Tz+\tilde A)=0$ and $u^Tz=0$, i.e. $z=-\s\l\tilde
A$. Then
\begin{eqnarray}\label{Gl:DasIstJaEinProjektor!}
-\l^T\s\l+{1}_E\;=\;P_{\ker\l},
\end{eqnarray}

\noindent where ${1}_E$ is the $E\times E$ unit-matrix and $P_{\ker
\l}$ is the orthogonal projector onto the subspace
$\ker\l\subset\C^E$. In particular
\begin{eqnarray}
\l^Tz+\tilde A\;=\;P_{\ker\l}\tilde A.
\end{eqnarray}
\end{Lemma}

\noindent\textbf{Proof:} Since
\begin{eqnarray}\label{Gl:OrhtogonaleZerlegung}
\ker\l\,\oplus\,\img\l^T\;=\;1_E,
\end{eqnarray}

\noindent The statement (\ref{Gl:DasIstJaEinProjektor!}) can be
rephrased as follows:
\begin{eqnarray}
\l^T\s\l\;=\;P_{\img\l^T},
\end{eqnarray}

\noindent which is the projector onto the image of $\l^T$. Let
$v\in\img\l^T$, so $v=\l^T w$ for some $w\in\C^V$. Even more, since
$\l^Tu=0$, we even can choose $w$ to be orthogonal to $u$:
$w\in\{u\}^{\perp}$. Then
\begin{eqnarray*}
\l^T\s\l\,v\;=\;\l^T\s(\l\l^T)w\;=\;\l^Tw\;=\;v,
\end{eqnarray*}

\noindent since by definition $\s$ is the inverse of $\l\l^T$ on
$\{u\}^{\perp}$.

Let, on the other hand, $v\in\{\img\l^T\}^{\perp}=\ker\l$. Then
\begin{eqnarray*}
\l^T\s\l\,v\;=\;0
\end{eqnarray*}

\noindent trivially. Thus, $\l^T\s\l$ leaves vectors in $\img\l^T$
invariant and annihilates vectors from the orthogonal complement of
$\img\l^T$. Hence $\l^T\s\l\;=\;P_{\img\l^T}$, from which it follows
that
\begin{eqnarray*}
-\l^T\s\l+{1}_E\;=\;P_{\ker\l}.
\end{eqnarray*}

\noindent This was the first claim, the second one
\begin{eqnarray*}
\l^Tz+\tilde A\;=\;P_{\ker\l}\tilde A.
\end{eqnarray*}

\noindent follows immediately.\\

\noindent The Lemmas \ref{Lem:LoesungenVonGleichungssystemen} and
\ref{Lem:DasIstJaEinProjektor!} enable us to rewrite
(\ref{Gl:AllesMitVektorenAusgedrueckt}) as
\begin{eqnarray}\label{Gl:JetztKoennenWirIntegrieren!}
(\ref{Gl:AusgangsFormel})\;&=&\;V\sqrt{\frac{2\pi}{t}}^E\int_{\R^{V}}\frac{d\tilde\phi_1\ldots
d\tilde\phi_{V}}{(2\pi)^{V-1}}\;\d\left(u^T\tilde\phi\right)\\[5pt]\nonumber
&&\qquad\qquad\qquad\times
\sum_{m_{V},\ldots,m_E\in\Z}\;\exp\left(-\frac{\tilde\phi^T\l\l^T\tilde\phi}{2t}
\,-\,\frac{\tilde A^TP_{\ker \l}\tilde A}{2t}\right).
\end{eqnarray}

\noindent We can now finally evaluate the gauge integrals in
(\ref{Gl:JetztKoennenWirIntegrieren!}) with the help of Kirchhoff's
theorem. Since the delta-function in the integrand of
(\ref{Gl:JetztKoennenWirIntegrieren!}) assures that we only
integrate over the orthogonal complement of $u$, instead of $\R^V$,
and Kirchhoff's theorem \ref{Thm:Kirchhoff} assures that the
Kirchhoff-matrix $\l\l^T$ is positive definite there, we can
immediately evaluate the integral:
\begin{eqnarray}
\int_{\R^V}\frac{d\tilde\phi_1\cdots
d\tilde\phi_V}{(2\pi)^{V-1}}\,\d\left(u^T\tilde\phi\right)\;\exp\left(-\frac{\tilde\phi^T\l\l^T\tilde\phi}{2t}\right)\;
&=&\;\sqrt\frac{t}{2\pi}^{V-1}\;\frac{1}{\sqrt{\prod_{a=2}^V\m_a}}\\[5pt]\nonumber
&=&\;\frac{1}{\sqrt{G\,V}}\;\sqrt\frac{t}{2\pi}^{V-1}
\end{eqnarray}

\noindent where $\m_2,\ldots,\m_V$ are the nonzero eigenvalues of
the Kirchhoff-matrix, and $G$ is the number of different possible
maximal trees in the graph $\g$. With this, the gauge-invariant
coherent state can be written as
\begin{eqnarray*}
(\ref{Gl:AusgangsFormel})\;=\;\sqrt\frac{V}{G}\sqrt\frac{2\pi}{t}^{E-V+1}\sum_{m_V,\ldots,m_E\in\Z}\exp\left(-\frac{(A-2\pi
m)^TP_{\ker\l}(A-2\pi m)}{2t}\right)
\end{eqnarray*}

\noindent where $A=z-\phi$ is the vector containing $A_k=z_k-\phi_k$
in its $k$-th component, and $m$ being the vector containing $0$ in
the first $V-1$ components and $m_V,\ldots,m_E$ in the last $E-V+1$
components.

As already stated, the kernel of $\l$ is $E-V+1$-dimensional. Let
$l_1,\ldots,l_{E-V+1}$ be an orthonormal basis of
$\ker\l\subset\R^E$. Define
\begin{eqnarray}\label{Gl:EichinvarianteKombinationen}
z_{\n}^{gi}\;:=\;l_{\n}^Tz,\qquad
\phi^{gi}_{\n}\;:=\;l_{\n}^T\phi,\qquad m^{gi}_{\n}\;:=\;l_{\n}^Tm.
\end{eqnarray}

\noindent With this and
$P_{\ker\l}=\sum_{\n=1}^{E-V+1}l_{\n}l_{\n}^T$, we get our final
formula
\begin{eqnarray}\label{Gl:FinaleFormel}
\Psi_{[z]}^t(\phi)&&\\[5pt]\nonumber
\;&=&\;\sqrt\frac{V}{G}\sqrt\frac{2\pi}{t}^{E-V+1}\sum_{m_V,\ldots,m_E\in\Z}\exp\left(-\sum_{\n=1}^{E-V+1}\frac{(z^{gi}_{\n}-\phi_{\n}^{gi}-2\pi
m^{gi}_{\n})^2}{2t}\right).
\end{eqnarray}

\noindent The gauge-invariant coherent state only depends on the
$z^{gi}_{\n}$ and $\phi^{gi}_{\n}$, which are gauge-invariant
combinations of the $z_k$ and $\phi_k$. That the linear combinations
(\ref{Gl:EichinvarianteKombinationen}) are gauge-invariant, is clear
from the construction, but one can immediately see this from the
following: Perform a gauge-transformation, which shifts the $\phi_k$
by $\l_{ka}\tilde\phi_a$. So, in matrices, one has
$\phi\to\phi+\l^T\tilde\phi$. Thus,
\begin{eqnarray*}
\phi^{gi}_{\n}\;=\;l^T_{\n}\phi\;\longrightarrow\;l^T_{\n}(\phi+\l^T\tilde\phi)\;=\;l^T_{\n}\phi\;+\;l^T_{\n}\l^T\tilde\phi\;=\;l^T_{\n}\phi\;=\;\phi^{gi}_{\n},
\end{eqnarray*}

\noindent where $l_{\n}\in\ker\l$ has been used, from which it
follows that $\l l_{\n}=0$, so $l_{\n}^T\l^T=0$. Thus, the linear
combinations of $\phi$ in $\phi^{gi}_{\n}$ are all gauge-invariant.
The same holds true, of course, for the $z_{\n}^{gi}$ and
$m_{\n}^{gi}$. So, the coherent states depend only on
gauge-invariant combinations of $\phi$, which was clear from the
beginning, but can now be seen explicitly. Note that the basis
$\{l_{\n}\}_{\n=1}^{N-V+1}$ is, of course, not unique, but can be
replaced by any other basis $l'_{\n}=R_{\n\m}l_{\m}$ with $R\in
O(E-V+1)$.\\

Compare the formula for the gauge-invariant coherent state
(\ref{Gl:FinaleFormel}) with the formula for the gauge-variant
coherent states on $E$ edges
(\ref{Gl:EichvarianterKohaerenterZustandAufNemGraphen}). Up to a
factor of $(V/G)^{1/2}$, the similarity is striking. One could be
led to the conclusion that gauge-invariant coherent states are
nothing but gauge-variant coherent states, only depending on
gauge-invariant quantities. The fact that the gauge-invariant
configuration space is diffeomorphic to $U(1)^{E-V+1}$, supports
this guess.

However, this is not true. The reason is that the summation
variables $m_V,\ldots,m_E$ are placed in wrong linear combinations
in the formula. In particular, a gauge-invariant state is \emph{not}
\begin{eqnarray}\nonumber
\Psi_{[z]}^t(\phi)\;&\neq&\;\sqrt\frac{V}{G}\sqrt\frac{2\pi}{t}^{E-V+1}\sum_{m^{gi}_1,\ldots,m^{gi}_{E-V+1}\in\Z}
\exp\left(-\sum_{\n=1}^{E-V+1}\frac{(z^{gi}_{\n}-\phi_{\n}^{gi}-2\pi
m^{gi}_{\n})^2}{2t}\right)\\[5pt]\label{Gl:SchoenWaers!}
&=&\;\sqrt\frac{V}{G}\psi^t_{z^{gi}}(\phi^{gi}).
\end{eqnarray}

\noindent Of course, from the form (\ref{Gl:FinaleFormel}) one
cannot deduce a priori that the $m^{gi}_{\n}$ could not, probably,
be reordered in a way, maybe by an intelligent choice of $l_{\n}$
and/or suitable shifting of summations, such that a form like
(\ref{Gl:SchoenWaers!}), possibly with different $t$ for different
variables, could be possible. But already at simple examples like
the $3$-bridge graph show that this cannot be done. It could be, if
one is lucky (in particular, on the $2$-bridge graph), but
generically a gauge-invariant coherent state is no complexifier
coherent state depending on gauge-invariant variables.

\subsection{Peakedness of gauge-invariant coherent states}

\noindent In this chapter, we will shortly investigate the
peakedness properties of the gauge-invariant coherent states. In
particular, we will show that they are peaked on gauge-invariant
quantities. Let $\g$ be a graph with $E$ edges. Then, a complexifier
coherent state is then labeled by $E$ complex numbers
$z_1,\ldots,z_E$ and a semiclassicality parameter $t>0$. Such a
state is given by
\begin{eqnarray}
\psi^t_z(\phi)\;=\;\sqrt\frac{2\pi}{t}^E\sum_{m_1,\ldots,m_E\in\Z}\exp\left(-\sum_{k=1}^E\frac{(z_k-\phi_k-2\pi
m_k)^2}{2t}\right).
\end{eqnarray}

\noindent The corresponding gauge-invariant coherent states,
obtained by applying the projector onto the gauge-invariant
sub-Hilbert-space, has, in the last section, been shown to be
\begin{eqnarray*}
\Psi^t_{[z]}(\phi)\;=\;\sqrt{\frac{V}{G}}\sqrt\frac{2\pi}{t}^{E-V+1}\sum_{m_V,\ldots,m_E\in\Z}\exp\left(-\sum_{\n=1}^{E-V+1}\frac{(z^{gi}_{\n}-\phi^{gi}_{\n}-2\pi
m_{\n}^{gi})^2}{2t}\right).
\end{eqnarray*}

\noindent Here $G$ is the number of different possible maximal trees
is the graph $\g$ and $\phi_{\n}^{gi}=l_{\n}^T\phi$, where
$l_1,\ldots,l_{E-V+1}$ is an orthonormal base for the kernel
$\ker\l\subset \R^E$ of the incidence matrix $\l$ of $\g$. Also,
$z_{\n}^{gi}=l_{\n}^Tz$ and $m_{\n}^{gi}=l_{\n}^Tm$, where $m$ is
the vector containing zeros in the first $V-1$ entries, and $m_V$ to
$m_E$ in the last $E-V+1$ entries.\\

\noindent The inner product between two gauge-invariant coherent
states $\Psi^t_{[w]}$ and $\Psi^t_{[z]}$ is, as one can easily
calculate, given by
\begin{eqnarray}\nonumber
\left\langle\Psi^t_{[w]}\Big|\Psi^t_{[z]}\right\rangle\;=\;\sqrt\frac{V}{G}\sqrt\frac{\pi}{t}^{E-V+1}
\sum_{m_V,\ldots,m_E\in\Z}\exp\left(-\sum_{\n=1}^{E-V+1}\frac{(\bar
w^{gi}_{\n}-z^{gi}_{\n}-2\pi m_{\n}^{gi})^2}{t}\right).\\[5pt]\label{Gl:EichinvariantesInneresProdukt}
\end{eqnarray}

\noindent With $z_k=\phi_k-ip_k$, i.e. by splitting the phase-space
points into configuration- and momentum variables, one immediately
gets a formula for the norm of a gauge-invariant coherent state:

\begin{eqnarray}\nonumber
\left\|\Psi^t_{[z]}\right\|^2\;=\;\sqrt\frac{V}{G}\sqrt\frac{\pi}{t}^{E-V+1}
\sum_{m_V,\ldots,m_E\in\Z}\exp\left(4\sum_{\n=1}^{E-V+1}\frac{(p^{gi}_{\n}-\pi
i m_{\n}^{gi})^2}{t}\right).\\[5pt]\label{Gl:NormOFAGaugeInvariantCoherentState}
\end{eqnarray}

\noindent Note that there is, apart from $m=0$, no combination of
$m_V,\ldots,m_E$ such that the corresponding $m_{\n}^{gi}=0$ for all
$\n=1,\ldots,E-V+1$. If there is one such combination, there are
infinitely many of these combinations, hence infinitely many equally
large terms. So, if there were, then the sum in
(\ref{Gl:EichinvariantesInneresProdukt}) would not exist at all. But
we know that the sum in (\ref{Gl:EichinvariantesInneresProdukt}) is
absolutely convergent, so there is no such combination.

What we just said is equivalent to saying that
\begin{eqnarray*}
P_{\text{ker}\,\l}\left(\begin{array}{c}0\\\vdots\\0\\m_V\\\vdots\\
m_E\end{array}\right)\;\neq\;0\qquad\mbox{for all }m_V,\ldots
m_E\in\Z,
\end{eqnarray*}

\noindent which is, of course, due to the fact that the last $E-V+1$
components correspond, by construction, to the gauge-invariant
directions on $U(1)^E$. In the limit of $t\to 0$, the norm of a
gauge-invariant coherent state
(\ref{Gl:NormOFAGaugeInvariantCoherentState}) can be written as

\begin{eqnarray}
\left\|\Psi^t_{[z]}\right\|^2\;&\leq&\;\sqrt\frac{V}{G}\sqrt\frac{\pi}{t}^{E-V+1}
\sum_{m_V,\ldots,m_E\in\Z}\exp\left(4\sum_{\n=1}^{E-V+1}\frac{(p^{gi}_{\n})^2-\pi^2
(m_{\n}^{gi})^2}{t}\right)\\[5pt]\nonumber
&=&\;\sqrt\frac{V}{G}\sqrt\frac{\pi}{t}^{E-V+1}\,\exp\left({4\sum_{\n=1}^{E-V+1}\frac{(p_{\n}^{gi})^2}{t}}\right)
\sum_{m_V,\ldots,m_E\in\Z}\;\exp\left(-4\pi^2\sum_{\n=1}^{E-V+1}\frac{
m^TP_{\text
{ker}\l}m}{t}\right)
\end{eqnarray}

\noindent Define
\begin{eqnarray*}
K\;:=\;\min_{\|m\|=1}\left\|P_{\text {ker}}m\right\|\;>\;0.
\end{eqnarray*}

\noindent With this, $m^TP_{\text {ker}}m\;\geq\,K^2\|m\|^2$, so we
get
\begin{eqnarray}\nonumber
\sum_{m_V,\ldots,m_E\in\Z}\;\exp\left(-4\pi^2\sum_{\n=1}^{E-V+1}\frac{
m^TP_{\text
{ker}\l}m}{t}\right)\;&\leq&\;\sum_{m_V,\ldots,m_E\in\Z}\exp\left(-4\pi^2K^2\frac{\|m\|^2}{t}\right)\\[5pt]\label{Gl:RechnungWarumManDieMsWeglassenKann}
&=&\;\left[\sum_{n\in\Z}\exp\left(\frac{-4\pi^2K^2}{t}n^2\right)\right]^{E-V+1}\\[5pt]\nonumber
&=&\;1\,+\,O(t^{\infty}).
\end{eqnarray}

\noindent Thus, we can write
\begin{eqnarray}\label{Gl:NormDerEichinvariantenZustaende}
\left\|\Psi^t_{[z]}\right\|^2\;=\;\sqrt\frac{V}{G}\sqrt\frac{\pi}{t}^{E-V+1}
\sum_{m_V,\ldots,m_E\in\Z}\exp\left(4\sum_{\n=1}^{E-V+1}\frac{(p^{gi}_{\n})^2}{t}\right)(1+O(t^{\infty})).
\end{eqnarray}

\noindent For the inner product between complexifier coherent
states, one has
\begin{eqnarray}\label{Gl:ShiftenDerArgumenteDerKomplexifiziererZustaende}
\left\langle\psi_w^t\Big|\psi^t_z\right\rangle\;=\;\left\langle\psi_0^t\Big|\psi^t_{z-\bar
w}\right\rangle,
\end{eqnarray}

\noindent as can be readily deduced from the explicit formula of the
inner product between two complexifier coherent states. This is also
true for the gauge-invariant coherent states, which have
\begin{eqnarray}
\left\langle\Psi_{[w]}^t\Big|\Psi^t_{[z]}\right\rangle\;=\;\left\langle\Psi_{[0]}^t\Big|\Psi^t_{[z-\bar
w]}\right\rangle.
\end{eqnarray}

\noindent This can either be deduced by applying the gauge-projector
onto (\ref{Gl:ShiftenDerArgumenteDerKomplexifiziererZustaende}), or
directly from formula (\ref{Gl:EichinvariantesInneresProdukt}).

So, in order to show that the overlap of two gauge-invariant
coherent states, labeled by $[z]$ and $[w]$, is peaked at $[z]=[w]$,
one only has to show that the overlap between a state labeled by
$[z]$ and $\Psi^t_{[0]}$ is peaked at $[z]=[0]$. With
(\ref{Gl:NormDerEichinvariantenZustaende}) and $z=\phi-ip$, we get

\begin{eqnarray*}
\frac{\left\langle\Psi_{[0]}^t\Big|\Psi^t_{[z]}\right\rangle}{\left\|\Psi^t_{[0]}\right\|\;\left\|\Psi^t_{[z]}\right\|}
\;&=&\;\sum_{m_V,\ldots,m_E\in\Z}\exp\left(-\sum_{\n=1}^{E-V+1}\frac{(\phi^{gi}_{\n}-ip^{gi}{\n}+2\pi
m_{\n}^{gi})^2}{t}\,-\,\sum_{\n=1}^{E-V+1}\frac{2(p^{gi}_{\n})^2}{t}\right)\\[5pt]
&&\qquad\times(1+O(t^{\infty}))\\[5pt]
&=&\;\sum_{m_V,\ldots,m_E\in\Z}\exp\left(-\sum_{\n=1}^{E-V+1}\frac{(\phi^{gi}_{\n}-2\pi
m_{\n}^{gi})^2+(p_{\n}^{gi})^2}{t}+2i\frac{p_{\n}^{gi}(\phi^{gi}_{\n}-2\pi
m_{\n}^{gi})}{t}\right)\\[5pt]
&&\qquad\times(1+O(t^{\infty})).
\end{eqnarray*}

\noindent If the $\phi^{gi}_{\n}$ are close to zero, then the term
with all $m^{gi}_{\n}=0$, which corresponds to all $m_k=0$, is
significantly larger than the other terms. So this can, with similar
arguments as in (\ref{Gl:RechnungWarumManDieMsWeglassenKann}),  be
further simplified into
\begin{eqnarray}
\frac{\left\langle\Psi_{[0]}^t\Big|\Psi^t_{[z]}\right\rangle}{\left\|\Psi^t_{[0]}\right\|\;\left\|\Psi^t_{[z]}\right\|}
\;&=&\;\exp\left(-\sum_{\n=1}^{E-V+1}\frac{(\phi^{gi}_{\n})^2+(p_{\n}^{gi})^2}{t}+2i\frac{p_{\n}^{gi}\phi^{gi}_{\n}}{t}\right)(1+O(t^{\infty})).
\end{eqnarray}

\noindent This approaches $1$ if the gauge-invariant quantities
$\phi^{gi}$ and $p^{gi}$ are close to zero, but as soon as the
gauge-invariant quantities are away from zero, the expression
becomes tiny, due to the tiny $t$. It follows that the overlap is
peaked at gauge-invariant quantities.\\

\section{Summary and conclusion}

\noindent This is the first of two  articles concerning the
gauge-invariant coherent states for Loop Quantum Gravity. In this
one, we have replaced the gauge-group $G=SU(2)$ of LQG by the much
simpler $G=U(1)$, the case $G=U(1)^3$, which is also of interest for
LQG, follows immediately. We have investigated the gauge-invariant
coherent states, in particular we have computed their explicit form
and their overlap. The results found are very encouraging: While the
complexifier coherent states are peaked on points in the kinematical
phase space, which contains gauge information, the gauge-invariant
coherent states, which are labeled by gauge-equivalence classes, are
also sharply peaked on these. In particular, the overlap between two
gauge-invariant coherent states labeled with different gauge orbits
tends to zero exponentially fast as the semiclassicality parameter
$t$ tends to zero. Even more, it could be shown that the overlap is
actually a Gaussian in the gauge-invariant variables.

This shows the good semiclassical properties of these states: As $t$
tends to zero, different states become approximately orthogonal very
quickly, suppressing the quantum fluctuations between them. Also,
the expectation values of operators corresponding to gauge-invariant
kinematical observables (such as volume or area) are approximated
well, which immediately follows from the corresponding properties of
the gauge-variant CCS states.

This shows that the gauge-invariant coherent states are in fact
useful for the semiclassical analysis of the gauge-invariant sector
of LQG, and is the first step on the road to \emph{physical}
coherent states.

Apart from the nice semiclassical properties, the computation has
revealed an explicit connection between the gauge-invariant sector and the
graph topology. In particular, the formula for the gauge-invariant
coherent states on a graph $\g$ contains the incidence matrix $\l$
of $\g$. In contrast, the CCS are simply a product of states on each
edge of the graph, hence have no notion of which edges are connected
to each other and which are not, while the gauge-invariant coherent
states explicitly contain information about the graph topology. This
is simply due to the fact that the set of gauge-invariant degrees of
freedom depend on the graph topology and can be computed by
graph-theoretic methods.\\

While the results for $G=U(1)$ are quite encouraging, the case of
ultimate interest for LQG is $G=SU(2)$, which is much more
complicated. We will address this topic in the following article,
which will deal with this issue and try to establish as much results
as possible from $U(1)$, where the problem could be solved
completely and analytically, also for $SU(2)$.

\section*{Acknowledgements}

BB would like to thank Hendryk Pfeiffer for the discussions about
gauge-invariant functions and cohomology. Research at the Perimeter
Institute for Theoretical Physics is supported by the Government of
Canada through NSERC and by the Province of Ontario.

\appendix

\section{Cohomology with values in abelian
groups}\label{Ch:Gauge-invariantFunctionsOfU(1)}

\noindent In the following we will write down the definition for
singular cohomology with values in an abelian group. This will allow
for a compact notation of the gauge-invariant Hilbert space. In
particular, we will characterize the cohomology spaces in question
to arrive at a better understanding what to expect, when computing
the gauge-invariant coherent states on graphs and their overlaps.
Note that we will employ, for brevity, the notation

\begin{eqnarray}
A^B\;:=\;\{f:B\to A\text{ any map}\}
\end{eqnarray}

\noindent for the set of maps from any set $B$ to any set $A$.

 Consider a CW complex $K$, i.e. a topological space that
is successively built up of $n$-cells (n-dimensional closed balls),
such that the intersection of two cells is a collection of
lower-dimensional sub-cells, and around each point there is a
neigbourhood that contains finitely many cells.
In particular, any graph in $\Sigma$ is a CW complex of dimension 1,
i.e. consisting only of 1-cells (the edges), that intersect at the
0-cells (the vertices).

 Let $K^n$ be the set of all $n$-cells in the CW complex
$K$. Let $G$ be an abelian group, then define $C^n(K, G)$ to be the
set $G^{K^n}$, i.e. the set of all maps from $K^n$ to $G$. Then
$C^n(K,G)$ is obviously an abelian group, simply by defining the
group multiplication pointwise. This group is obviously homomorphic
to $G^{|K^n|}$.

We then define a chain by
\begin{eqnarray}\label{Gl:Kettenkomplex}
\{1\}\;\stackrel{\d}{\longrightarrow}\;C^0(K,G)\;\stackrel{\d}{\longrightarrow}\;C^1(K,G)\;
\stackrel{\d}{\longrightarrow}\;C^2(K,G)\;\stackrel{\d}{\longrightarrow}\;\ldots
\end{eqnarray}

\noindent where $\d: C^n(K,G)\to C^{n+1}(K,G)$ is defined by the
following rule: Let $f:K^n\to G $ be an element of $C^n(K,G)$. Then,
for an $n+1$-cell $c$ we define
\begin{eqnarray}\label{Gl:DefinitionBoundaryOperator}
\d f(c)\;:=\;f(v_1)^{\s_1}\cdot\ldots\cdot
f(v_{|K^n|})^{\s_{|K^n|}},
\end{eqnarray}

\noindent where $v_1,\ldots v_{|K^n|}$ are all $n$-cells and the
factors $\sigma_k$ are defined to be $+1$ if $v_k$ is part of the
boundary of $c$ and the orientation of $v_k$ is the same as the
induced one from $c$, $-1$ if $v_k$ is in the boundary of $c$ but
the induced orientation from $c$ and the given one on $v_k$ differ
by a sign, $0$ if $v_k$ is not part of the boundary of $c$.

Note that $\d$ is a group homomorphism, which follows from the
abeliness of $G$. Hence, for each $n$, both sets
$\ker\;\d:C^n(K,G)\to C^{n+1}(K,G)$ and
$\text{img}\,\d:C^{n-1}(K,G)\to C^n(K,G)$ are subgroups of
$C^n(K,G)$, where the kernel of a group homomorphism is defined to
be the set of all elements being mapped to the unit element.

One can explicitly check that with this definition, that the map
\begin{eqnarray*}
\d\d:C^n(K,G)\;\longrightarrow\;C^{n+2}(K,G)
\end{eqnarray*}

\noindent maps every $C^n(K,G)$ to the unit element in
$C^{n+2}(K,G)$. It follows that even $\text{img}\,\d:C^{n-1}(K,G)\to
C^n(K,G)$ is a subgroup of $\ker\;\d:C^n(K,G)\to C^{n+1}(K,G)$.
Thus, one can define the quotients
\begin{eqnarray*}
H^n(K,G)\;:=\;\frac{\ker\;\d:C^n(K,G)\to
C^{n+1}(K,G)}{\text{img}\,\d:C^{n-1}(K,G)\to C^n(K,G)},
\end{eqnarray*}

\noindent which is called the \emph{$n$-th cohomology group of $K$
with values in $G$}. As the name suggests, this is of course also an
abelian group.\\

\noindent The definition above is fairly general, but we will now
see what it means for the specific case of the CW complex being an
oriented graph $\g$ (with the orientations of the vertices all being
set to the number $+1$). We keep the abelian group $G$ arbitrary for
the moment, having in mind the application to $G=U(1)$ or $G=U(1)^3$
lateron.\\


Let us consider a graph $\g$, consisting of a set of edges $E(\g)$
and vertices $V(\g)$. The chain in (\ref{Gl:Kettenkomplex}) is then
simply
\begin{eqnarray*}
\{1\}\;\stackrel{\d}{\longmapsto}\;G^{V(\g)}\;\stackrel{\d}{\longmapsto}\;G^{E(\g)}\;\stackrel{\d}{\longmapsto};\{1\}.
\end{eqnarray*}

\noindent The only nontrivial map is $\d:G^{V(\g)}\to G^{E(\g)}$.
For every edge $e\in E(\g)$, $b(e)$ and $f(e)$ are called the
beginning- and endpoint of the edge, and are by construction both
elements of $V(\g)$. So let $k:V(\g)\to G$ be an element of
$G^{V(\g)}$. Then the definition of $\d$ given above implies
\begin{eqnarray}\label{Gl:ActionOfCoboundaryOperator}
(\d  k)_e\;=\;k_{b(e)}\,k_{f(e)}^{-1},
\end{eqnarray}

\noindent so $\d k $ is a map from $E(\g)$ to $G$, that is an
element of $G^{E(\g)}$. The only nontrivial cohomology groups we can
form are then
\begin{eqnarray}\label{Gl:DefinitionZerothCohomology}
H^0(\g,G)\;&=&\;\frac{\ker\;\d:G^{V(\g)}\to
G^{E(\g)}}{\text{img}\,\d:\{1\}\to
G^{V(\g)}}\;=\;\ker\;\d:G^{V(\g)}\to
G^{E(\g)},\\[5pt]\label{Gl:DefinitionFirstCohomology}
H^1(\g,G)\;&=&\;\frac{\ker\;\d:G^{E(\g)}\to
\{1\}}{\text{img}\,\d:G^{V(\g)}\to
G^{E(\g)}}\;=\;\frac{G^{E(\g)}}{\text{img}\,\d:G^{V(\g)}\to
G^{E(\g)}}.
\end{eqnarray}

\noindent These two groups have nice interpretations in terms of the
graph topology, which are stated by the following lemma:

\begin{Lemma}\label{Lem:TopologyInterpretationOfCohomologyGroups}
Let $\g$ be a graph (connected, oriented, finitely many edges,
embedded in a 3-manifold $\Sigma$). Then, for any abelian group $G$,
we have
\begin{eqnarray}\label{Gl:MeaningOfCohomologyGroups}
H^0(\g,G)\;&\simeq&\;G,\\ H^1(\g, G)\;&\simeq&\;\text{Hom
}(\pi_1(\g),G),
\end{eqnarray}

\noindent Loosely speaking, $H^0(\g,G)$ counts the connected parts
of $\g$, and $H^1(\g, G)$ counts the numbers of "holes" in $\g$.
\end{Lemma}

\noindent\textbf{Proof:} The proof is quite standard, but we will
still repeat it here.

By (\ref{Gl:ActionOfCoboundaryOperator}) and
(\ref{Gl:DefinitionZerothCohomology}), we see that $H^0(\g,G)$
consists of all maps $k$ from $V(\g)$ to $G$, such that, for every
edge $e$, $k_{b(e)}\,k_{f(e)}^{-1}=1$. Since the graph is connected,
this is equivalent to saying that the map $k$ assigns to each vertex
$v\in V(\g)$ the same element in $G$:
\begin{eqnarray*}
k_v\;=\; h \quad\mbox{for some } h\in G\mbox{ and all }v\in V(\g).
\end{eqnarray*}

\noindent The group of all such maps is then clearly equivalent to
the group $G$ itself, since the graph $\g$ is connected. So we have
\begin{eqnarray*}
H^0(\g,G)\;\simeq\;G.
\end{eqnarray*}

\noindent To show the second part of
(\ref{Gl:MeaningOfCohomologyGroups}), consider a maximal tree $\t$
in the graph $\g$. A maximal tree is a subgraph such that each
vertex of $\g$ is also contained in $\t$ (i.e. $V(\g)=V(\t)$), and
the graph $\t$ contains no closed loops. Call all edges in $\g$ that
are not in $\t$ leaves. Maximal trees exist for all graphs, although
they are far from unique. The number of leaves in a graph, though,
is independent from the choice of $\t$.

To compute $H^1(\g,G)$, we have to compute the orbits of the
subgroup $\d(G^{V(\g)})\subset G^{E(\g)}$. We do this by showing
that, to each $h\in G^{E(\g)}$, we can apply an element of
$\d(G^{V(\g)}$, such that the result is an element $\tilde h\in
G^{E(\g)}$ such that $\tilde h_e=1$ for all $e\in E(\t)$. In short,
we show that one can gauge fix the group elements on the edges
belonging to the tree $\t$ to 1. The remaining distribution of
elements $\tilde h_e$ for leaves $e$ is unique, due to the fact that
the group $G$ is abelian.

Consider an element $h$ of $G^{E(\g)}=C^1(\g,G)$, i.e. a
distribution $(h_{e_1},\ldots,h_{e_E})$ of elements in $G$ among the
edges in $E(\g)$. Construct an element $k\in V(\g)$ by the following
method: Choose a vertex $v$ in $V(\g)$. For each other vertex $w\in
V(\g)$, there is a unique path from $w$ to $v$ along edges in $\t$,
since $\t$ contains no loops. So, in order to get from $w$ to $v$,
one has to go, say, along edges $e_1,\ldots e_n$, either parallel or
antiparallel to the orientation of the $e_i$. Define the $k_w$ to be
the product
\begin{eqnarray}\label{Gl:GaugeFixingProcedure}
k_w\;=\;h_{e_1}^{\pm 1}h_{e_{2}}^{\pm 1}\cdots h_{e_n}^{\pm 1},
\end{eqnarray}

\noindent where the element $h_{e_i}$ is contained in the product,
if the path from $w$ to $v$ is parallel to the orientation of $e_i$.
If the path is antiparallel, then take $h_{e_i}^{-1}$ instead.

Thus, an element $k\in G^{V(\g)}$ is defined. Now consider the
product $\tilde h:=\d k\cdot h$. It is quite easy to see that the
element $\tilde h$ assigns $1\in G$ to each $e\in E(\t)$: consider
an $e\in E(\t)$. The path from $f(e)$ to $v$ passes through $b(e)$,
or the other way round. Assume the first to be the case, the other
case works analogously. We have then
\begin{eqnarray*}
k_{f(e)}\;=\;h_e\,k_{b(e)},
\end{eqnarray*}

\noindent since the path from $f(e)$ goes against the orientation of
$e$ to $b(e)$, and then is identical to the way from $b(e)$ to $v$,
since $\t$ contains no loops. So
\begin{eqnarray*}
\tilde h_e\;=\;k_{b(e)}\;h_e\;k_{f(e)}^{-1}\;=\;1.
\end{eqnarray*}

\noindent Thus, we have shown, the orbit of each element $h\in
G^{E(\g)}$ under the action of the subgroup $\d(G^{V(\g)})$ contains
an element $\tilde h$ such that only the elements assigned to the
leaves in $\g$ are potentially different from $1\in G$. One can also
see that the only element in $\d(G^{V(\g)})$ that leaves the
distribution of $1$ along the edges of $E(\t)$ unchanged, is an
element $k\in \ker \d$, i.e. such that $k_v=h$ for some $h\in G$ and
all $v\in V(\g)$. The multiplication with $\d k$ leaves $G^{E(\g)}$
invariant, since $G$ is abelian.
We thus see that the element of $\tilde h$ is unique for each $h\in
E(\g)$, hence does not depend on the choice of the vertex $v$. This
shows that each orbit in $G^{E(\g)}$ under the action of
$\d(G^{V(\g)})$ determines uniquely a distribution of group elements
in $G$ among the leaves of $\g$.\vspace{10pt}

\noindent Since $\t$ contains no loops, it is contractible. Consider
the flower graph $\tilde \g$ that one obtains by contracting the
tree $\t$ to a point. This graph contains just one vertex $V$ and a
number of edges, all starting and ending at $V$, corresponding to
the number of leaves of $\g$. Note that $H^1(\tilde \g,G)=E(\tilde
\g)^G$. From this and our considerations above it follows that there
is a natural group isomorphism between $H^1(\g,G)\simeq
H^1(\tilde\g,G)$. It is clear that the first fundamental group
$\pi_1(\tilde \g)$ is freely generated by the elements of
$E(\tilde\g)$. In particular, $H^1(\tilde\g,G)\simeq\text{
Hom}(\pi_1(\tilde\g),G)$.

Since $\t$ contains no loops, the tree is contractible, hence
$\tilde \g$ is a retraction of $\g$. In particular, both graphs are
homotopy equivalent. Since the first fundamental group is a homotopy
invariant, we conclude
\begin{eqnarray*}
H^1(\g,G)\;\simeq\;\text{Hom}(\pi_1(\g),G).
\end{eqnarray*}

\noindent In particular, $H^1(\g,G)\simeq G^L$, where $L$ is the
number of leaves in $\g$ (which is independent of the choice of the
maximal tree $\t$).

\subsection{Gauge-invariant functions}

\noindent The notion of gauge-invariant cylindrical functions fits
nicely into the framework of cohomology. Remember that a
gauge-variant function on a graph $\g$ is determined via
(\ref{Gl:CorrespondingFunction}) by a function of a number of copies
of the gauge group $G$:
\begin{eqnarray*}
\tilde f:\underbrace{G\times\cdots\times G}_{\mbox{one for each edge
in }E(\g)}\;\to\;\C
\end{eqnarray*}

\noindent that is square-integrable with respect to the product
Haar-measure $d\m_H^{\otimes|E(\g)|}$. These function constitute the
Hilbert space $\H_{\g}$, and with the notions of the previous
sections, we identify this space to be
\begin{eqnarray}
\H_{\g}\;\simeq\;L^2\big(C^1(\g,G),\;d\m_H^{\otimes|E(\g)|}\big).
\end{eqnarray}

\noindent The gauge transformed $\tilde f$ is determined by letting
the gauge group $G$ act on every vertex $v\in V(\g)$ via
(\ref{Gl:ActionOfGaugeGroup}):
\begin{eqnarray*}
\a_{k_{v_1},\ldots,k_{v_V}}\tilde
f\;\big(h_{e_1},\ldots,h_{e_E}\big)\;=\;\tilde
f\big(k_{b(e_1)}^{-1}h_{e_1}
k_{f(e_1)},\;\ldots\;,k_{b(e_E)}^{-1}h_{e_E} k_{f(e_E)}\big),
\end{eqnarray*}

\noindent where $b(e)$ and $f(e)$ are the vertices sitting at the
beginning and the end of the edge $e$ respectively. \\

Not only do we recognize the gauge transformation group as the space
$G^{V(\g)}=C^0(\g,G)$ from the previous section, one can see readily
the connection between the gauge transformation $\a$ and the
coboundary operator $\d$:
\begin{eqnarray*}
(\a_{g_1,\ldots,g_V}\tilde f)(h_1,\ldots,h_E)\;=\;\tilde
f\big(\d(g_1,\ldots,g_V)\cdot (h_1,\ldots,h_E)\big),
\end{eqnarray*}

\noindent where $\cdot$ means group multiplication in
$C^1(\g,G)=G^{E(\g)}$. So, the gauge-invariant functions on the
graph $\g$ are just the functions on the group $G^{E(\g)}$ that are
invariant under the action of $\d(G^{V(\g)})$. We conclude that the
gauge-invariant functions coincide with the functions on the first
cohomology class

\begin{eqnarray}
\P\H_{\g}\;\simeq\;L^2\big(H^1(\g,G),\;d\m\big),
\end{eqnarray}

\noindent where the measure $d\m$ is the quotient measure of
$d\m_H^{\otimes|E(\g)|}$ under the action of the gauge
transformation group $G^{V(\g)}$, which, since $H^1(\g,G)$ is a
group for abelian $G$, can be identified with the normalized Haar
measure on $H^1(\g,G)$.
%
%


\begin{thebibliography}{}


\bibitem{INTRO} Thiemann, T.: \emph{Introduction to modern canonical quantum general
relativity} (Cambridge Monographs on Mathematical Physics) Cambridge University Press 2006 


\bibitem{ROVELLISBUCH} Rovelli, C.: \emph{Quantum Gravity} (Cambridge Monographs on Mathematical Physics) Cambridge University Press 2004

\bibitem{INTRO3} Smolin, L.: \emph{An invitation to Loop Quantum
Gravity} [arXiv:hep-th/0408048]

\bibitem{ALLMT} Ashtekar, A., Lewandowski, J., Marolf, D., Mourao, J.,
Thiemann, T. \emph{Quantization of diffeomorphism invariant theories
of connections with local degrees of freedom} 1995 J. Math. Phys.
{\bf 36} 6456 [arXiv:gr-qc/9504018]

\bibitem{LOST} Lewandowski, J., Okolow, A., Sahlmann, H., Thiemann,
T. \emph{Uniqueness of diffeomorphism invariant states on
holonomy-flux algebras} 2006 Commun. Math. Phys. {\bf 267} No. 3,
703 [arXiv:gr-qc/0504147]

\bibitem{HENN-TEITEL} Henneuax, M., Teitelboim, C.: \emph{Quantization of gauge
systems} 1992 Princeton University Press

\bibitem{KECK}
Baranger, M.,   de Aguiar, M. A. M.,   Keck.,F.,  Korsch, H. J.,
Schellhaa{\ss}  B. \emph{Semiclassical approximations in phase space
with coherent states} 2001  J. Phys. A: Math. Gen.  {\bf 34} 7227;
(see also ibd 2002 {\bf 35} 9493;   2003 {\bf 36}   9795)

\bibitem{KORSCHCHAOS1} B. Mirbach, H. J. Korsch:
\emph{Phase Space Entropy and Global Phase Space Structures of
(Chaotic) Quantum Systems} 1995 Phys. Rev. Lett. \textbf{75}, 362

\bibitem{KORSCHCHAOS2} H. Wiescher and H. J. Korsch:
\emph{Intrinsic ordering of quasienergy states for mixed
regular/chaotic quantum systems: zeros of the Husimi distribution}
1997 J. Phys. A \textbf{30}, 1763

\bibitem{KLAUDER}  Klauder, J. R., Skagerstam, B. S.: \emph{Coherent states:
applications in physics and mathematical physics} 1985 Singapore:
World Scientific

\bibitem{VAN-VLECK} Van Vleck, J. H.  \emph{The Correspondence Principle in the Statistical Interpretation of Quantum Mechanics} 1928 Proc. Natl. Acad. Sci. USA
{\bf 14} 178

\bibitem{GLAUBER}  Glauber, R. J. \emph{Coherent and Incoherent States of the Radiation Field}  1963 Phys. Rev. {\bf 131}, 2766

\bibitem{HALL1}
Hall, B.  \emph{The Segal-Bargmann "coherent state'' transform for
compact Lie groups} 1994 J. Funct. Anal.  {\bf 122}       103,
\emph{The inverse Segal-Bargmann transform for compact Lie groups}
1997 J. Funct. Anal.  {\bf 143}       98

\bibitem{HALL3}
Hall, B. \emph{Phase space bounds for quantum mechanics on a compact
Lie group}  1997 Commun. Math. Phys.  {\bf 184}       233

\bibitem{KASTRUP} Kastrup., H.  \emph{Quantization of the canonically conjugate pair angle and orbital angular momentum}   2006 Phys. Rev. A {\bf 73},    052104
[arXiv:quant-ph/0510234]

\bibitem{KRP}
Kowalski, K.,   Rembieli\'nski, J.,  Papaloucas., L. C.
\emph{Coherent states for a quantum particle on a circle} 1996 J.
Phys. A: Math. Gen.  {\bf 29},    4149  [quant-ph/9801029]

\bibitem{CCS} Thiemann, T. \emph{Complexifier coherent states for quantum general relativity} 2001 Class. Quant. Grav. {\bf
18}, 2025 [arXiv:gr-qc/0206037]

\bibitem{GCS1}
Thiemann, T.  \emph{Gauge Field Theory Coherent States (GCS) I.
General Properties} 2001  Class.Quant.Grav. \textbf{18}, 2025
[arXiv:hep-th/0005233]

\bibitem{GCS2} Thiemann, T., Winkler, O. \emph{Gauge field theory coherent states (GCS) II. Peakedness properties} 2001 Class. Quant. Grav. {\bf
18}, 2561 [arXiv:hep-th/0005237],  \emph{Gauge Field Theory Coherent
States (GCS) : III. Ehrenfest Theorems}  2001 Class. Quant. Grav.
{\bf 18}, 4629 [arXiv:hep-th/0005234]

\bibitem{TINA1} Giesel, K., Thiemann, T.\emph{Algebraic Quantum
Gravity (AQG) I. Conceptual Setup}, 2007 Class. Quant. Grav. {\bf
24}, 2465 [gr-qc/0607099], \emph{Algebraic Quantum Gravity (AQG) II.
Semiclassical Analysis}, 2007 Class. Quant. Grav. {\bf 24}, 2499
[gr-qc/0607100], \emph{Algebraic Quantum Gravity (AQG) III.
Semiclassical Perturbation Theory}, 2007 Class. Quant. Grav. {\bf
24}, 2565 [gr-qc/0607101]

\bibitem{SNF} Rovelli, C., Smolin, L. \emph{Spin Networks and Quantum
Gravity} 1995 Phys.Rev. D {\bf 52} 5743 [arXiv:gr-qc/9505006]

\bibitem{VARA} Varadarajan, M. \emph{The graviton vacuum as a distributional state in kinematic Loop Quantum Gravity} 2005 Class. Quant. Grav. {\bf 22}
1207-1238 [arXiv:gr-qc/0410120]

\bibitem{QFTCST} Sahlmann, H., Thiemann, T. \emph{Towards the QFT on Curved Spacetime Limit of QGR. I: A General Scheme}
2006 Class. Quant. Grav. {\bf 23}, 867 [arXiv:gr-qc/0207030],
 \emph{Towards the QFT on Curved
Spacetime Limit of QGR. II: A Concrete Implementation} 2006 Class.
Quant. Grav. {\bf 23}, 909 [arXiv:gr-qc/0207031]

\bibitem{GICS-II} Bahr, B., Thiemann, T. \emph{Gauge-invariant coherent states for Loop Quantum Gravity II: Non-abelian gauge
groups} [arXiv:gr-qc/0709.4636]

\bibitem{QSD1} Thiemann, T. \emph{Quantum Spin Dynamics (QSD)} 1998 Class. Quant. Grav. {\bf 15}, 839 [arXiv:gr-qc/9606089],
\emph{Quantum Spin Dynamics (QSD) II}, 1998 Class. Quant. Grav. {\bf
15}, 875 [arXiv:gr-qc/9606090], \emph{QSD III : Quantum Constraint
Algebra and Physical Scalar Product in Quantum General Relativity}
1998 Class. Quant. Grav. {\bf 15} , 1207  [arXiv:gr-qc/9705017]

\bibitem{PHOENIX} Thiemann, T. \emph{The Phoenix Project: Master Constraint Programme for Loop Quantum Gravity} 2006 Class. Quant. Grav. \textbf{23}, 2211  [arXiv:gr-qc/0305080]

\bibitem  {QSD8} Thiemann, T. \emph{Quantum Spin Dynamics VIII. The Master Constraint} 2006 Class. Quant. Grav. {\bf
23},  2249 [arXiv:gr-qc/0510011]

\bibitem{CE} Flori, C., Thiemann, T. \emph{Semiclassical analysis of the Loop Quantum Gravity volume operator utilising cylindrically consistent complexifier coherent
states}, in preparation

\bibitem{GRAPH} Biggs., N.: \emph{Algebraic Graph Theory} 2nd Ed.
1993 Cambridge Mathematical Library

















\end{thebibliography}
\end{document}